\PassOptionsToPackage{sort&compress}{natbib}
\documentclass[final,3p,times]{elsarticle}
\usepackage{booktabs}
\usepackage{ulem}
\usepackage{float}
\usepackage{siunitx} 
\usepackage{natbib}
\usepackage{color}
\usepackage{tabularray}
\usepackage[version=4]{mhchem}
\usepackage{amssymb}
\usepackage{amsmath}
\begin{document}

%% -------------------- Title --------------------
\title{Time-Resolved Pore-Scale Imaging of Multiphase Dissolution during \ce{CO2}-Saturated Brine Injection into a Carbonate: Competition between Hydrocarbon Mobilisation and Swelling}

%% -------------------- Authors --------------------
\author[label1,label2]{Qianqian Ma}
\author[label2]{Rukuan Chai}
\author[label1,label2]{Zhuangzhuang Ma}
\author[label1,label2]{Yanghua Wang}
\author[label2]{Martin J. Blunt}
\author[label2]{Branko Bijeljic\corref{cor1}}

\cortext[cor1]{Corresponding author.}
\ead{b.bijeljic@imperial.ac.uk}

%% -------------------- Affiliations --------------------
\affiliation[label1]{%
  organization = {Resource Geophysics Academy, Imperial College London},
  addressline  = {London, SW7 2BP},
  country      = {United Kingdom}}

\affiliation[label2]{%
  organization = {Department of Earth Science and Engineering, Imperial College London},
  addressline  = {London, SW7 2AZ},
  country      = {United Kingdom}}

\begin{frontmatter}
%% Abstract
\begin{abstract}
 We present time-resolved pore-scale experiments in which \ce{CO2}-saturated brine was injected into a water-wet Ketton limestone sample containing residual hydrocarbon under reservoir conditions (8~MPa, 50~\textdegree C) and monitored by 4D X-ray microtomography. Equivalent pore-network models were extracted at each scan time to track pore geometry, topology, and fluid occupancy, while fluid--fluid and fluid--rock interfacial areas and the effective reaction rate were determined from segmented images. The dissolution rate is non-monotonic in time and proceeds through three successive regimes, consistent with a shifting balance between hydrocarbon swelling and ganglion mobilisation, which together control advective access to reactive surfaces. In the initial advection-dominated regime, pore-throat widening leads to ganglia mobilisation and efficient acidic brine delivery to reactive surfaces. The second, dissolution-inhibited regime is marked by up to two orders of magnitude reduction in effective reaction rate. Pore-network analysis indicates that swollen hydrocarbon ganglia persistently occupy the largest throats throughout this regime. This occupancy is associated with a reorganisation of the advective flow field into preferential flow paths and stagnant zones. We interpret the rate suppression as primarily reflecting a path-dependent loss of advective access to reactive surfaces, with subordinate contributions from localised \ce{H+} depletion near ganglia and reduced near-wall mass transfer in widened flow paths. The inhibited state persists until hydrocarbon is displaced from the largest throats, after which, in the third stage, advective access improves and rock dissolution accelerates. These results show that the effective dissolution rate in residual-hydrocarbon-bearing carbonate depends dynamically on the competition between hydrocarbon swelling and ganglion mobilisation, governing advective access to mineral surfaces.
\end{abstract}

%% Keywords

\end{frontmatter}

\section*{Plain Language Summary}
When carbon dioxide (CO\textsubscript{2}) is injected underground for long-term storage, it dissolves into the resident brine and forms a weak acid that slowly eats away at carbonate rock. Many target storage sites are depleted hydrocarbon reservoirs that still contain small, trapped pockets of residual hydrocarbon. We used time-lapse X-ray imaging to watch, at the scale of individual pores, what happens when CO\textsubscript{2}-enriched brine floods a limestone sample containing trapped oil.

Dissolution was not steady. It proceeded rapidly at first, then nearly stopped for several hours before recovering. The pause coincided with CO\textsubscript{2} migrating from the brine into the trapped hydrocarbon phase, causing it to swell and block the widest flow channels in the rock. This rerouted the acidic brine away from most of the mineral surface, starving the rock of the acid needed to dissolve it. Once the swollen trapped phase was eventually displaced, dissolution resumed. These findings show that residual hydrocarbon can reversibly slow rock dissolution---an effect that current models of CO\textsubscript{2} storage do not capture, but that matters for predicting how injection changes rock structure over time.

\section*{Highlights}
\begin{itemize}
\item Time-resolved micro-CT captures three stages of carbonate dissolution during \ce{CO2}-saturated brine injection in hydrocarbon-bearing rock.
\item Dissolution is initially rapid and then drops a hundredfold as hydrocarbon swells and blocks access of acidic brine to the mineral surface.
\item Dissolution accelerates once hydrocarbon is mobilised.
\item The effective dissolution rate reflects a competition between hydrocarbon swelling and ganglion mobilisation, which together govern advective access to mineral surfaces.
\end{itemize}

\section{Introduction}
\label{sec1}
Predicting storage efficiency and injectivity in geological carbon sequestration (GCS) requires an understanding of pore-scale reactive transport and multiphase flow processes in carbonate formations\cite{krevor2023subsurface, szulczewski2012lifetime}---yet these mechanisms remain incompletely understood, creating fundamental uncertainties that propagate directly to formation-scale predictions \cite{chai2025pore}.
During \ce{CO2} injection, dissolution into resident brine forms carbonic acid that lowers pore-fluid pH, initiating reactive dissolution of carbonate minerals.
This geochemical alteration continuously modifies porosity, permeability, and capillary pressure---quantities that directly govern injectivity, plume migration, and long-term storage integrity \cite{chai2025multiphase}.
Most pore-scale studies, however, have focused on idealized brine--\ce{CO2}--carbonate systems, leaving the role of residual hydrocarbons largely unaddressed.
Depleted hydrocarbon reservoirs represent particularly attractive \ce{CO2} storage sites owing to their proven structural integrity and existing subsurface infrastructure. They inevitably retain residual hydrocarbons that introduce a second fluid phase and fundamentally alter reactive transport dynamics \cite{zhou2024review}.
Accurate prediction of reactive transport in such multiphase systems is therefore essential---not only for assessing storage security, but also for quantifying how geochemical pore-network alteration propagates to formation-scale permeability and fluid connectivity \cite{wang2025porea}.
However, the role of a residual hydrocarbon phase in modifying reactive transport in carbonates remains poorly characterised under realistic GCS conditions.

Pore-scale investigations of \ce{CO2} saturated brine--carbonate systems under single-phase conditions have established a coherent mechanistic framework for reactive dissolution.
The competition among advection, diffusion, and surface reaction--- parameterized by the Péclet (\textit{Pe}) and Damköhler (\textit{Da}) numbers---governs the transition between uniform dissolution and channel formation \cite{li2022three, zhou2022pore, xie2026pore, noirielInvestigationPorosityPermeability2004,noirielHydraulicPropertiesMicrogeometry2005, luquotExperimentalDeterminationPorosity2009, smith2013evaporite, ottWormholeFormationCompact2015, luhmann2014experimental}.
Concurrent micro-CT studies have quantified porosity, permeability, and effective reaction rates for these regimes: the average reaction rate is consistently at least an order of magnitude lower than the batch reaction rate, owing to mass-transfer limitations imposed by preferential channelling \cite{menkeDynamicThreeDimensionalPoreScale2015, menkeReservoirConditionImaging2016},
and the spatial distribution of minerals \cite{al-khulaifiPoreScaleDissolution2019, sabo2021porosity}.
In parallel, two-phase \ce{CO2}--brine systems have been studied from a
mass-transfer and trapping perspective \cite{chai2025multiphase, wang2025pore}. Pore-scale simulations demonstrate that the \ce{CO2}--brine interfacial area, rather than the local mass-transfer coefficient, is the dominant control on dissolution under varying wettability conditions \citep{yang2023pore}; more broadly, quantifying the fluid-fluid interfacial area is recognized as essential for characterizing reactive multiphase transport in porous media \citep{gao2021two}.
Two-phase flow involving hydrocarbon and \ce{CO2} saturated brine has also been imaged in situ, revealing that the presence of hydrocarbon phase changes the dissolution pattern, permeability evolution and reaction rate \cite{ma2026pore, akindipePoreMatrixDissolution2022}.

During two-phase flow involving hydrocarbon and \ce{CO2}-saturated brine, the residual non-aqueous phase introduces coupled mass-transfer pathways: transport of acidic brine to mineral surfaces, which drives dissolution~\cite{al-khulaifiReactionRatesChemically2017a}, and interphase transfer of \ce{CO2} from brine into the residual hydrocarbon phase~\cite{bijeljic2003multicomponent, foroozesh2016mathematical, foroozesh2018physics}.
Because these processes can locally compete for dissolved \ce{CO2} and simultaneously modify fluid occupancy, three controls on the apparent dissolution rate are present.
First, \ce{CO2} partitioning into the residual phase may reduce aqueous \ce{CO2} concentration and brine acidity, thereby suppressing carbonate dissolution~\cite{mahzari2019integrated}.
Second, \ce{CO2} uptake may drive swelling and redistribution of the residual phase, changing capillary pressure, remobilising or stabilising trapped ganglia, and modifying the fluid--fluid interfacial area available for interphase \ce{CO2} transfer~\cite{sharbatian2018full}.
Third, persistent occupation of the most conductive throats or principal advective pathways by the trapped phase may reorganise the flow network and reduce the effective access of acidic brine to mineral surfaces outside these pathways~\cite{jimenez2020homogenization}.
The relative importance of these chemical, interfacial, and geometric-hydrodynamic controls remains unresolved by direct pore-scale observation in carbonate systems and is still poorly constrained by the existing literature~\cite{li2022effect}.

To address this gap, we performed time-resolved X-ray microtomography core-flooding experiments in which \ce{CO2}-saturated brine was injected into a water-wet Ketton limestone sample containing residual hydrocarbon at reservoir conditions (8~MPa, 50~\textdegree C). Time-resolved micro-CT was used to capture dynamic changes in pore structure, porosity, fluid occupancy and interfacial area. The pressure drop across the sample was continuously monitored. Based on these 4D datasets, equivalent pore-network models were extracted at every scan time to track the geometric and topological evolution of the system. Fluid--fluid and fluid--rock interfacial areas and the effective reaction rates were quantified directly from the time-series images. The observables underlying this study are the 4D micro-CT time series and the synchronous pressure record; all subsequent analyses derive from these two data streams.

We seek to understand the history-dependent impact of fluid distribution and pore occupancy on dissolution dynamics. We hypothesize that, rather than being governed solely by instantaneous brine–rock interfacial area, pore velocity, or \textit{Pe}–\textit{Da} values, the effective reaction rate response depends on dynamically evolving multiphase phenomena which distinguish it as a unique signature of multiphase dissolution. To this end, we present direct pore-scale observations of the mass transfer dependent \ce{CO2} dissolution in the oil phase, oil swelling and displacement, and fluid-solid reaction, providing an image-based analysis of the manifestations of  these underlying mechanisms.

\section{Materials and Methods}
\subsection{Materials}
\label{subesc1}
A cylindrical sample of oolithic Ketton limestone (\SI{12}{\milli\meter} in length and \SI{6}{\milli\meter} in diameter) was studied. The rock was composed of \SI{99.1}{\percent} calcite and characterized by a well-connected, bimodal pore size distribution, with micro-pore and macro-pore throat radius peaks below 0.1~\si{\micro\meter} and above 10~\si{\micro\meter}, respectively, separated by almost three orders of magnitude~\cite{patmonoaji2025differential}. The rock had a helium porosity of 0.239 ± 0.008 ~\cite{patmonoaji2025differential}. The absolute permeability of the sample was measured based on Darcy’s law at three flow rates under single-phase (brine) flow conditions to be 1.95 ± 0.03 Darcy.

Decane was selected as the oil phase \cite{akindipePoreMatrixDissolution2022}.
Ketton limestone is generally regarded as strongly water-wet \cite{singh2016imaging}. The synthetic brine consisted of 5~wt\% \ce{NaCl} and 1~wt\% \ce{KCl} dissolved in deionized water, doped with 30~wt\% potassium iodide (\ce{KI}) to enhance X-ray contrast during imaging \cite{lin2021drainage}.
At the experimental conditions of \SI{50}{\celsius} and \SI{8}{\mega\pascal}, the viscosities of brine and decane were $0.82$~mPa.s \cite{patmonoaji2025differential} and $0.838$~mPa.s (provided by PubChem, open chemistry database), respectively.

\subsection{Experimental Methods}
\label{subsec2}

\textbf{i. Preparation of CO\textsubscript{2}-saturated brine.}
The brine was equilibrated with CO\textsubscript{2} in a high-pressure reactor at 8~MPa and 50~\textdegree C for one week.

\begin{figure}[H]
  \centering
  \includegraphics[width=0.85\textwidth]{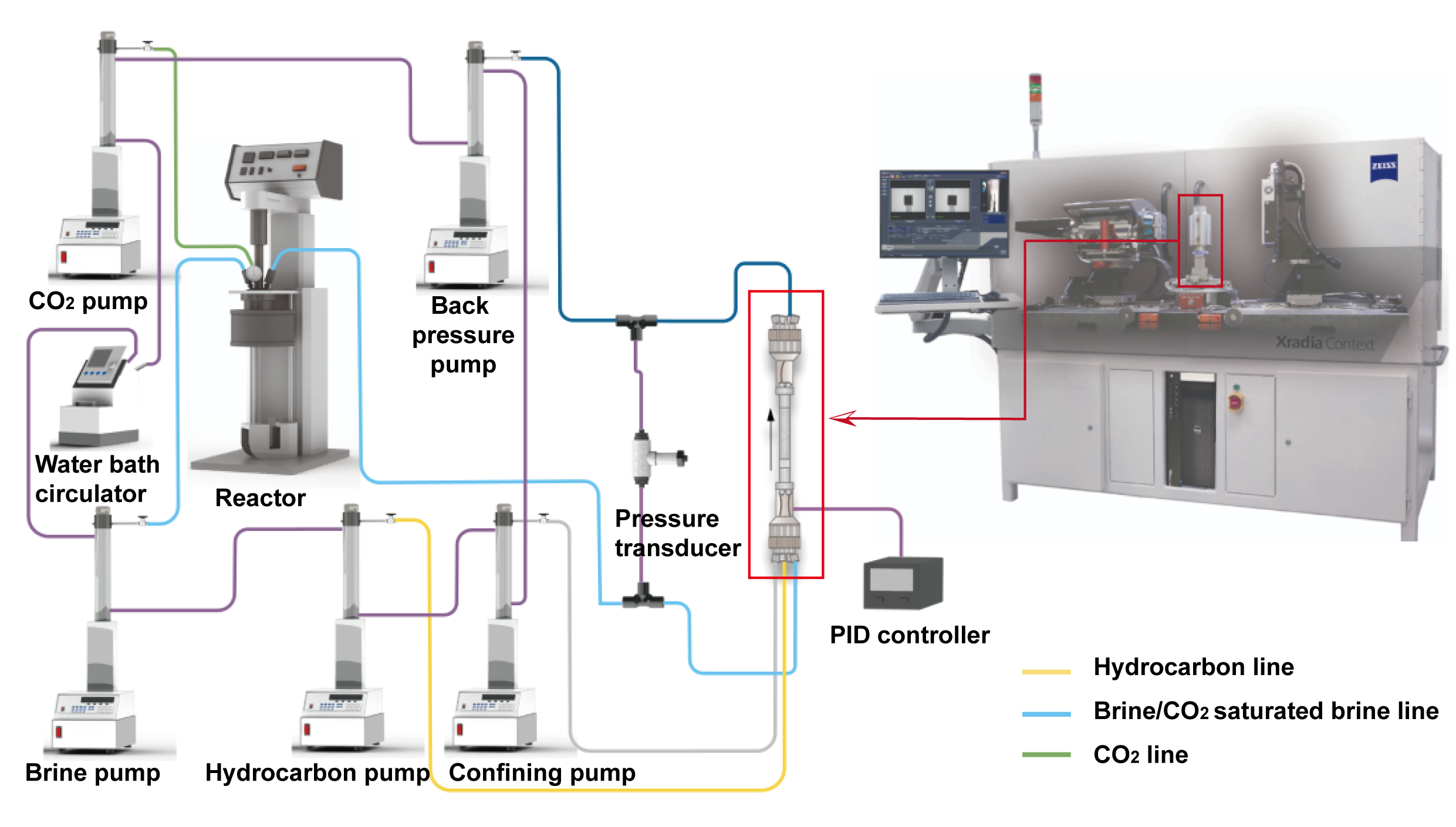}
  \caption{The experimental apparatus including the flow loop and micro-CT scanner.}
  \label{fig:1}
\end{figure}

\textbf{ii. System assembly.}
The Ketton limestone sample was enclosed in a Viton sleeve and mounted into a core holder. The assembled core holder was then installed in a CT scanner (Versa XRM-500 X-ray Microscope) and connected to the fluid delivery and receiving system, pressure transducer, and other experimental components, as shown in Figure~\ref{fig:1}.

\textbf{iii. System cleaning.}
 A confining pressure of 2~MPa was applied to ensure proper sealing and eliminate bypass flow between the sample and sleeve. The temperature was then increased to 50$^\circ$C. CO$_2$ was injected for 30~minutes (approximately 740 pore volumes (PV)) to flush residual fluids and fines, followed by 24~hours of vacuuming to evacuate the remaining gases. 
 
\textbf{iv. Baseline imaging.}
A high-resolution dry scan was acquired to serve as a baseline for subsequent comparisons.

\textbf{v. Brine saturation.}
The sample was saturated with brine at a constant flow rate of 0.05~mL\,min$^{-1}$ for 100 pore volumes (PVs). The back pressure and confining pressure were then steadily increased to 8~MPa and 10~MPa, respectively, and the temperature was maintained at 50~$^\circ$C, after which a brine saturated scan was performed. Throughout the experiment, injection occurred through the bottom inlet.

\textbf{vi. Establishment of initial oil saturation (\(S_{\mathrm{oi}}\)).}
Decane was injected at 0.05~$\mathrm{mL\,min^{-1}}$ for 100 PVs to establish \(S_{\mathrm{oi}}\), after which a corresponding micro-CT scan was obtained.

\textbf{vii. Establishment of residual oil saturation (\(S_{\mathrm{or}}\)).}
Brine was re-injected at 0.05~$\mathrm{mL\,min^{-1}}$ for 20~PVs to achieve \(S_{\mathrm{or}}\), followed by another scan. The resulting \(S_{\mathrm{or}}\) value was 30.2\%.

\textbf{viii. Reactive flooding with CO\textsubscript{2}-saturated brine.}
(i) Injection commenced at 0.1~$\mathrm{mL\,min^{-1}}$ for an initial period 
    of 73~min.
(ii) The flow rate was then alternated cyclically: briefly elevated to 0.5~$\mathrm{mL\,min^{-1}}$ for ${\sim}$10~min to ensure periodic replenishment of fresh \ce{CO2}-saturated brine throughout the pore network, then reduced to 0.1~$\mathrm{mL\,min^{-1}}$ for ${\sim}$47~min, during which the imposed flow was held below the threshold required to mobilise trapped oil ganglia, allowing the multiphase configuration to evolve under its intrinsic \ce{CO2}-partitioning and swelling dynamics whilst also minimising motion-induced blurring of fluid--fluid and fluid--solid interfaces during imaging. This cycle was repeated throughout the remainder of the experiment.
(iii) The total injection duration was 646~min, comprising the 
    initial 73~min low-flow period followed by approximately 
    10 complete cycles of the ${\sim}$57~min (10 + ${\sim}$47~min) alternating flow sequence.  

\subsection{Image Acquisition}
\label{subsec:2.3}
Two distinct micro-CT imaging protocols were employed: a high-spatial-resolution protocol for the detailed static characterization of the initial pore structure, and a high-temporal-resolution protocol to monitor the dynamic evolution of the pore structure during reactive transport.
To characterize the evolution of fluid occupancy and pore structure, a series of high-resolution micro-CT scans were conducted throughout the experiment. Prior to flooding, a dry scan of the sample was acquired to capture the initial pore geometry. This was followed by a brine-saturated scan after full saturation and a scan after decane injection to determine the initial oil distribution. All these scans used 2601 projections with an energy of \SI{100}{\kilo\electronvolt} and a power of \SI{9}{\watt}. The scans were conducted at a voxel resolution of \SI{3.57}{\micro\meter} and required approximately \SI{12}{\hour} each to ensure optimal image quality and segmentation accuracy. This resolution resulted in a corresponding field of view (FOV) of \SI{7.14}{\milli\meter} \texttimes \SI{7.14}{\milli\meter} \texttimes \SI{7.14}{\milli\meter}. To obtain a sufficiently representative image size at this high resolution, we performed two consecutive scans along the sample's axial direction and subsequently computationally stitched the resulting image data.

For the time-resolved imaging, a baseline scan was first acquired after brine re-injection to determine the residual oil saturation. The dynamic imaging sequence commenced at the onset of CO$_2$-saturated brine injection (\(t = 0\)), during which 9 successful scans were acquired. These dynamic scans, along with the residual oil saturation scan, were performed using 901 projections at an energy of \SI{100}{\kilo\electronvolt} and a power of \SI{9}{\watt}. Each scan required approximately \SI{45}{\minute} and used a voxel size of \SI{6.09}{\micro\meter}, which resulted in a \SI{12.18}{\milli\meter} \texttimes \SI{12.18}{\milli\meter} \texttimes \SI{12.18}{\milli\meter} field-of-view sufficient to capture the entire region of interest in a single acquisition.

The first two scans acquired immediately after injection were rendered unusable due to instrument malfunction. An additional scan acquired between \(t = 198\) and \(t = 244\)\,min was severely blurred owing to vigorous and complex fluid displacement during this period, and was therefore also excluded from quantitative analysis. Consequently, the first usable dynamic scan spans \(t = 83\) to \(t = 130\)\,min, referred to hereafter as the \SI{130}{\minute} scan, with subsequent scans named according to their end time as detailed in Table~\ref{tab:pv_comparison}. The entire reaction experiment lasted for a total of 646\,\si{\minute}. This imaging protocol was chosen to balance temporal resolution with image quality, enabling the accurate tracking of dissolution fronts and structural evolution without introducing significant motion artifacts or radiation damage.

\begin{table}[H]
\centering
\caption{Time-resolved micro-CT imaging protocol. All scans were acquired
             during continuous injection of CO$_2$-saturated brine at
             $0.1\;\mathrm{mL\;min^{-1}}$. Cumulative pore volumes (PV) are
             referenced to the end of each acquisition window.}
\sisetup{table-format=4.0}
\begin{tabular}{
  l 
  c 
  S[table-format=4.0]
}
\toprule
\textbf{Scan} & \textbf{Time interval (Start--End)} & \textbf{Pore volumes} \\
& {(\si{\minute})} & {} \\
\midrule
0 min scan & \multicolumn{2}{c}{Residual oil saturation scan} \\
\midrule
\SI{130}{\minute} scan & t=83 to t=130   & {330} \\
\SI{187}{\minute} scan & t=140 to t=187 & {465} \\
\SI{300}{\minute} scan & t=254 to t=300 & {677} \\
\SI{356}{\minute} scan & t=210 to t=356 & {782} \\
\SI{412}{\minute} scan & t=366 to t=412 & {886} \\
\SI{471}{\minute} scan & t=425 to t=471 & {990} \\
\SI{527}{\minute} scan & t=481 to t=527 & {1094} \\
\SI{584}{\minute} scan & t=537 to t=584 & {1195} \\
\SI{646}{\minute} scan & t=597 to t=646 & {1294} \\
\bottomrule
\end{tabular}
\label{tab:pv_comparison}
\end{table}

\subsection{Image Processing and Analysis}
Three-dimensional tomograms were reconstructed using Zeiss Reconstructor software correcting for beam hardening and centre shift.
Image processing was conducted in commercial image analysis software (Avizo) using a standardised workflow to ensure cross-scan comparability. Initially, images were denoised using a combination of non-local means and anisotropic diffusion filters. This was followed by intensity normalisation to standardise greyscale ranges. For the high-quality \SI{3.57}{\micro\meter} scan, two individual sections were aligned and stitched to form a single contiguous sample. Finally, all scans were co-registered to a common reference and resampled to a resolution of \SI{6.09}{\micro\meter}. This systematic approach guaranteed consistency across all imaging stages, facilitating accurate segmentation and quantitative analysis.

Phase segmentation employed a hybrid workflow to robustly delineate three primary phases: rock matrix, brine, and oil. The pre-reaction dataset (0\,min scan) was segmented through a combination of differential imaging~\cite{chai2022formation, chai2025pore, Gao2017}, interactive thresholding, and watershed-based segmentation~\cite{Lin2016}. The resulting 0\,min segmentation was used as the training label for a 3D Swin UNETR-based segmentation model~\cite{hatamizadeh2021swin}. The trained model was then applied independently to each post-reaction scan to generate initial predictions, which were subsequently inspected and manually corrected in Avizo to obtain the final time-series segmentations. Figure~S4 compares greyscale images with the final segmentations at 0, 300 and 646~min, and Movie~S1 shows greyscale images and segmentations at every scan time.

\textbf{Pre-reaction dataset (0\,min scan).}
(i) The rock mask was extracted from the dry reference scan.
(ii) Oil was segmented directly from the images without differential imaging.
(iii) Brine was isolated by subtracting the dry scan from the brine- or oil-saturated scans, thereby identifying the pores occupied by the aqueous phase.

\textbf{Post-reaction datasets.}
The pre-reaction image and its corresponding segmentation served as training data for a neural network model, which was subsequently applied to the post-reaction greyscale images to generate initial predictions. These predictions were then inspected and manually corrected in Avizo to produce the final segmented outputs. All training and inference were performed on an Nvidia A100 GPU. We segmented the full time series three times (Table~S2). For quantities calculated from all three segmentations, including $R_\mathrm{eff}$, Figures~\ref{fig:2}--\ref{fig:4} show the mean and error bars spanning the minimum and maximum values.

\subsection{Pore-Scale Simulation}
\label{sec:2.5}
Following image segmentation, the lower region of the segmented micro-CT volume ($5.9 \times 5.9 \times 6.1 \ \mathrm{mm}^3$) was selected for pore-scale flow simulation in the brine phase. The modelling approach follows the method developed by  Bijeljic et al. \cite{bijeljicInsightsNonFickianSolute2013} and Raeini et al. \cite{RAEINI20125653}, implemented within the \texttt{OpenFOAM} framework. The solver employs the finite volume method to simultaneously solve the continuity and steady-state incompressible Navier–Stokes equations:
\begin{align}
\nabla \cdot \mathbf{u} &= 0 \\
\rho \left( \frac{\partial \mathbf{u}}{\partial t} + \mathbf{u} \cdot \nabla \mathbf{u} \right) &= -\nabla p + \mu \nabla^2 \mathbf{u}
\label{NS}
\end{align}
where $p$ is the pressure ($\mathrm{Pa}$), and $u$ is the velocity ($\mathrm{m\,s^{-1}}$), both obtained for each voxel of the image; $\mu$ is the fluid (brine) viscosity ($\mathrm{Pa\,s}$); $\rho$ is the fluid density ($\mathrm{kg\,m^{-3}}$). The flow rate ($\mathrm{m^3\,s^{-1}}$) is calculated as $Q = \int u_x \, dA_x$, where $A_x$ is the cross-sectional area of the image ($\mathrm{m^2}$) and $u_x$ is the velocity in the direction of overall flow ($\mathrm{m\,s^{-1}}$). The Darcy velocity ($\mathrm{m\,s^{-1}}$) is then calculated as $q = \frac{Q}{L_y L_z}$, where $L_y$ and $L_z$ are the lengths of the image ($\mathrm{m}$).
A fixed 1 Pa pressure drop was applied along the flow direction, while all solid boundaries were treated as no-slip walls. In Section 3.4, the reported streamlines specifically correspond to the flow of brine in the presence of residual oil. These values were determined by simulating the flow field within the connected, brine-saturated pore network, where the residual oil was assumed to remain immobile. 

Additionally, the network extraction code based on the maximal ball algorithm \citep{PhysRevE.80.036307} was used to quantify the geometric and topological properties of the pore space. In the maximal ball algorithm, spheres are generated in the void space of the segmented images to determine the positions and diameters of pores; throats are the restrictions between pores. In this way, the pore space is topologically represented as a network of pores connected by narrow throats. 

\subsection{Dimensionless numbers and reaction rates}
In this study, the dimensionless number, Péclet (Pe) and Damköhler (Da), are used to characterize and quantify the reactive transport.
The Péclet number quantifies the relative efficiency of solute mass transfer through advection compared to diffusion \cite{peclet1827traite}:
\begin{align}
   \text{Pe} = \frac{\text{advective transport rate}}{\text{diffusive transport rate}} = \frac{u_{\text{av}} L_c}{D_m}
   \label{Pe}
\end{align}
where $D_m$ is molecular diffusion coefficient in brine ($\mathrm{m^2\,s^{-1}}$), $u_{\text{av}}$ is the average pore velocity, which is the Darcy velocity from the experiment divided by the product of porosity and brine saturation, while $L_c$ is characteristic length ($m$), calculated by \cite{mostaghimi2012simulation}:
\begin{align}
    L_c = \frac{\pi}{S}
\end{align}
where the specific surface area S (m$^{-1}$), is the image surface area per unit volume at the beginning of the time period, calculated by $V_B/As$, where $V_B$ is bulk volume, and $As$ is the surface area from image analysis.

The Damköhler number quantifies the ratio between the timescales of a chemical reaction and the mass transfer \cite{lasaga1984chemical}:
\begin{align}
    \text{Da} = \frac{\text{reaction rate}}{\text{advective transport rate}} = \frac{L_c}{u_{\text{av}}} k
    \label{12}
\end{align}
where $k$ is the chemical reaction rate constant (s$^{-1}$), calculated by: $ k = \frac{\pi r}{nL}$.
r is the mineral reaction rate ($\mathrm{mol\,m^{-2}\,s^{-1}}$), L is the sample length ($m$), and $n$ is calculated by \cite{menkeDynamicThreeDimensionalPoreScale2015}:
\begin{align}
    n = \frac{\rho_{\text{mineral}} f_{\text{mineral}}}{M_{\text{mineral}}}
\end{align}
where $\rho_{\text{}}$ is the mineral density, $M_{\text{mineral}}$ is their molecular mass. Therefore, equation \ref{12} can be rewritten as follows\cite{al-khulaifiReservoirconditionPorescaleImaging2018}:
\begin{align}
    \text{Da} = \frac{\pi r_{\text{mineral}}}{u_{\text{av}} n}
    \label{Da}
\end{align}
where $r_{\text{}}$ is the non-transport limited reaction rate. 

The effective reaction rate ($r_{\text{eff}}$) of mineral is determined as \cite{al-khulaifiReactionRatesChemically2017a}:
\begin{align}
    r_\text{eff} = \frac{\rho_\text{mineral} (1 - \phi_\text{unresolved}) \Delta \phi_\text{CT}}{M_\text{mineral} S \Delta t}
    \label{eq:reaction rate}
\end{align}
where $\Delta t$ is the time between scans (s), $\Delta \phi_\text{CT}$ is the corresponding change in porosity and $S$ is the image surface area per unit volume (\(\text{m}^{-1}\)). We observe almost no change in greyscale values for solid voxels containing sub-resolution features, indicating no measurable change in porosity in these regions. The unresolved porosity ($\phi_\text{unresolved}$) is calculated based on data from a sister sample drilled from the same block \cite{patmonoaji2025differential, ma2026pore}.

\section{Results and Discussion}
We examine the macroscopic non-monotonic dissolution behaviour (Section~\ref{subsec:3.1}), the pore-scale geometric and fluid-occupancy changes that underlie it (Section~\ref{subsec:3.2}), and the evolution of the reactive interface and effective dissolution rate (Section~\ref{subsec:3.3}). On this basis, in Section~\ref{subsec:3.4} we propose a mechanistic interpretation constrained by the imaging evidence above. Together, these analyses explain why the dissolution rate is non-monotonic in time and link its geometric, hydrodynamic, and reactive signatures. Section~\ref{subsec:3.5} and Section~\ref{subsec:3.6} then place the findings within broader modelling implications and define the applicability of the present study.

\subsection{Non-monotonic Dissolution Trajectory}
\label{subsec:3.1}
Figure~\ref{fig:2} shows the temporal evolution of global porosity, remaining oil saturation ($S_{or}$), and absolute oil volume fraction during the injection of CO$_2$-saturated brine. The evolution is non-monotonic in time and can be divided into three stages based on the trends in these quantities.
     
\subsubsection{Stage~1: Rapid dissolution and ganglia mobilisation
(0--187\,min)}
\label{subsec:3.1.1}
During Stage~1 (0--187\,min), porosity increases approximately linearly with time from 0.15 to 0.27 (Figure~\ref{fig:2}a), while both $S_\mathrm{or}$ (Figure~\ref{fig:2}c) and absolute oil volume fraction (Figure~\ref{fig:2}d) decline sharply. The CT image sequence shows progressive carbonate dissolution and oil displacement during this interval (Figure~\ref{fig:CT}a), and the size distribution of oil ganglia shifts towards smaller volumes over this interval (Figure~\ref{fig:CT}b), consistent with preferential mobilisation of larger ganglia from the pore network.

Together, these observations are consistent with an advection-dominated regime in which fresh acidic brine enters the pore network and dissolves the carbonate rock; the P\'eclet number (Eq.~\ref{Pe}) at 0.1~$\mathrm{mL\,min^{-1}}$ remains well above unity across all three scans (375, 196, and 131 for the 0, 130, and 187~min scans, respectively; Appendix Figure~\ref{appendix:PeDa}), despite a progressive decline as dissolution enlarges the pore space. The Damköhler number Da (Eq.~\ref{Da}) is small throughout the experiment ($\mathrm{Da} \ll 1$; Appendix Figure~\ref{appendix:PeDa}). The combination of $\mathrm{Pe} \gg 1$ and $\mathrm{Da} \ll 1$ places the system in the uniform-dissolution regime of a single-phase homogeneous porous medium \cite{golfier2002ability}. As we show later, with oil present in a heterogeneous rock, a markedly different dissolution behaviour emerges.

As pore throats widen, the capillary entry pressure ($P_c \propto 1/r$) decreases, making previously trapped oil ganglia easier to mobilise. The removal of these ganglia can connect additional brine-fed advective pathways and increase brine supply to mineral surfaces that were previously poorly connected to the bulk flow, producing a positive feedback between dissolution, ganglion mobilisation, and brine access.

Phase-equilibrium calculations indicate a strong thermodynamic driving force for \ce{CO2} transfer from brine into the hydrocarbon phase: the equilibrium \ce{CO2} mole fraction is \(\sim 0.89\) in the decane-rich phase~\cite{lemmon2018nist, span1996new, lemmon2006short, kunz2012gerg}, compared with \(\sim 0.018\) in the brine~\cite{duan1992equation, duan2003improved, weisenberger1996estimation}, a roughly 50-fold contrast in mole fraction. Despite this driving force, the image-derived oil volume does not show a measurable net increase. We interpret the absence of a net swelling signal as a consequence of competing mobilisation: dissolution-driven throat widening removes ganglia faster than \ce{CO2} partitioning can produce a CT-resolvable increase in the volume of retained ganglia.
The injection-pressure record during Stage~1 (Figure~\ref{fig:21}, top panel) shows an initial transient from ${\sim}$2~kPa to ${\sim}$15~kPa as flow is established in the sample, followed by relatively low-amplitude fluctuations at the baseline flow rate of 0.1~mL\,min$^{-1}$. 

\subsubsection{Stage~2: Oil ganglia swelling and dissolution imbibition (187--471\,min)}
\label{subsec:3.1.2}
During Stage~2 (187--471\,min), porosity reaches a plateau (Figure~\ref{fig:2}a,b) and the oil volume fraction rebounds (Figure~\ref{fig:2}d), showing that dissolution has slowed substantially while the remaining oil phase consists of ganglia that are more difficult to displace and instead swelled in place.
The CT image sequence supports this interpretation (Figure~\ref{fig:CT}a). Although oil droplet swelling is subtle when comparing individual CT slices side by side, it becomes more discernible when sequential images are viewed as a continuous time-lapse sequence. 
The oil-ganglion volume distributions (Figure~\ref{fig:CT}b,c) also show a shift toward larger ganglion volumes relative to the end of Stage~1. 
Most retained ganglia show limited positional displacement while their segmented volumes increase, consistent with in situ swelling rather than continued large-scale oil mobilisation.
Volume tracking of three representative ganglia (Figure~\ref{fig:CT}d,e,f) shows varying degrees of expansion; Ganglion~B exhibits an order-of-magnitude volume increase, attributed to the coalescence of neighbouring ganglia as the expanding oil phase reconnects previously isolated clusters.

These observations indicate that at $\sim$187\,min the system crosses a critical threshold at which the local rate of ganglion swelling begins to exceed the rate of ganglion mobilisation. This change can generate a positive feedback: \ce{CO2} uptake drives swelling, swelling extends ganglion residence in the pore network, and prolonged residence permits further \ce{CO2} uptake.
The injection-pressure record provides independent evidence. During the 197--244\,min interval at 0.1~mL\,min$^{-1}$, the pressure signal exhibits large-amplitude oscillations (Figure~\ref{fig:21}, middle panel; peak-to-trough ${\sim}\,10$\,kPa), in contrast to the smooth pressure increase observed during Stage~1 under the same flow rate. These oscillations could reflect repeated capillary build-up and partial breakthrough events, or flow diversion as rapid oil swelling altered the local fluid configuration. We note that the CT scan acquired during 197--244\,min was unusable due to motion-induced artefacts; the pressure record partially compensates for this gap but cannot provide spatial information. The first usable scan at 300~min shows enlarged, coalesced oil ganglia, consistent with swelling outpacing ganglion mobilisation as inferred from the pressure response.
The mechanism by which swelling-induced ganglion trapping suppresses dissolution—including a quantitative reconciliation with mass-balance constraints and the role of advective network reorganisation—is examined in Section~\ref{subsec:3.4}.

\subsubsection{Stage~3: Dissolution recovery ({>}\,471\,min)}
\label{subsec:3.1.3}
\begin{figure}[H]
  \centering
  \includegraphics[width=1\textwidth]{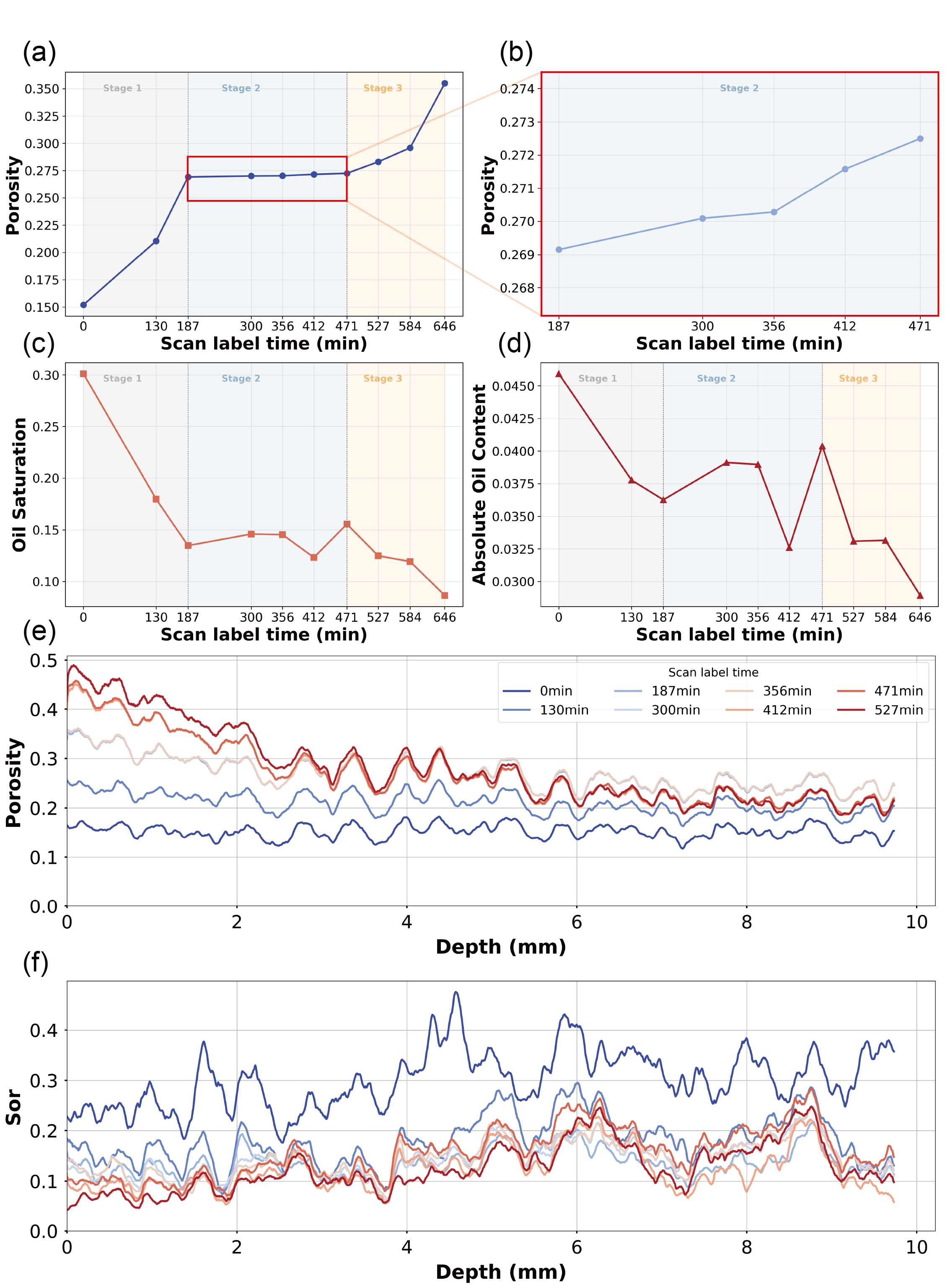}
  \caption{Temporal evolution of porosity, oil saturation, and 
oil content during reactive flooding. Each CT scan is labelled by the elapsed injection time at  which acquisition was completed; a complete scan required approximately \SI{47}{\minute}, with a \SI{10}{\minute} interval between consecutive scans. The \SI{0}{\minute}  label corresponds to the baseline scan completed before the  onset of reactive flooding. Post-reaction scans were  acquired sequentially: the first from $t = 83$ to \SI{130}{\minute} (labelled \SI{130}{\minute}), the second from $t = 140$ to \SI{187}{\minute} (labelled \SI{187}{\minute}), and so forth. 
All subsequent time references to CT data follow this scan-end convention. Vertical dashed lines delimit Stages~1--3.
\textbf{(a)}~Evolution of overall porosity across distinct 
reaction stages (Stages~1 to~3). 
\textbf{(b)}~Magnified view of porosity changes during the 
reaction-inhibited period (Stage~2). 
\textbf{(c)}~Evolution of average oil saturation. 
\textbf{(d)}~Evolution of absolute oil content (defined as 
the product of porosity and oil saturation, representing the 
oil volume fraction). 
\textbf{(e)}~Spatial profiles of local porosity along the 
core depth at selected scan label times. 
\textbf{(f)}~Corresponding spatial profiles of residual oil 
saturation (\(S_{or}\)) along the sample depth.}
  \label{fig:2}
\end{figure}

\begin{figure}[H]
  \centering
  \includegraphics[width=1\textwidth]{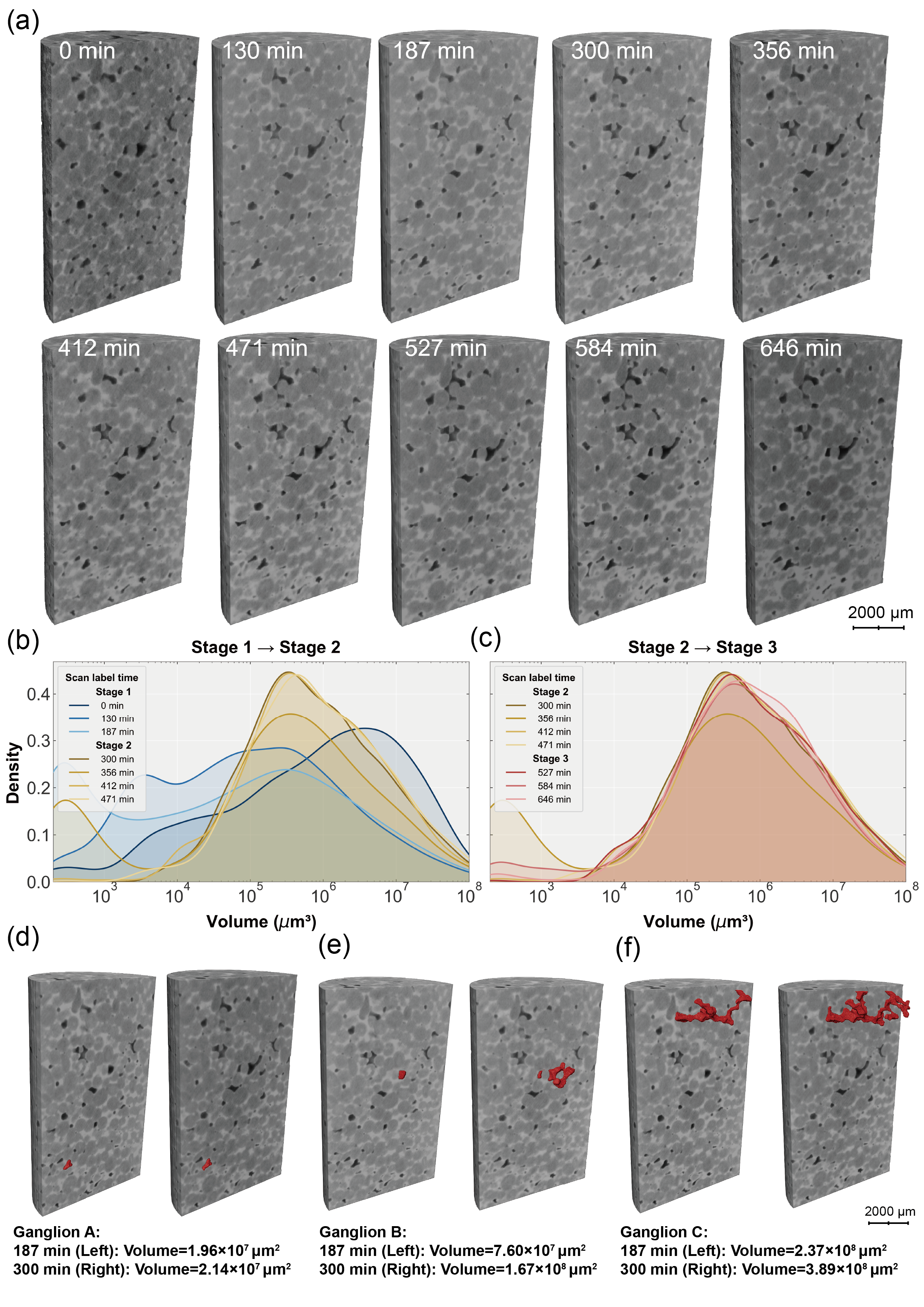}
  \caption{Greyscale image observations and oil ganglion evolution. (a)~Greyscale images depicting the temporal evolution of dissolution. Representative CT slices at successive time steps during Stage~2 (187--471\,min), showing the slowing of dissolution and swelling of trapped oil droplets. In these images, grey represents solids, black represents oil, and white represents brine. (b)--(c)~Oil ganglia distributions at different stages: the transition from Stage~1 to Stage~2 is marked by an overall increase in ganglia volume. (d)--(f)~Volume changes of three typical oil ganglia over time; oil ganglia are rendered in red.  Notably, Ganglion~B exhibits an order-of-magnitude volume increase, attributed to the coalescence of neighbouring ganglia as the expanding oil phase reconnects previously isolated clusters.}
  \label{fig:CT}
\end{figure}

During Stage~3 (471--646\,min), the image-derived porosity resumes increasing after the Stage~2 plateau (Figure~\ref{fig:2}a,b), whereas the absolute oil volume fraction declines (Figure~\ref{fig:2}d). This reversal indicates that carbonate dissolution becomes effective again and that the previously swollen retained oil phase was removed from the pore space. The CT image sequence supports this interpretation (Figure~\ref{fig:CT}a): renewed pore enlargement is accompanied by progressive reduction and reconfiguration of oil ganglia, consistent with renewed mobilisation of the swollen oil phase.
Because the injection conditions remained identical to those in Stage~2, the resumption cannot be attributed to any change in external acid supply; the continuous pressure record confirms uninterrupted fluid delivery throughout the experiment. Instead, the recovery appears to originate within the pore system itself. 
The mechanism underlying this recovery, including the role of throat reopening and the restoration of advective access, is presented in Section~\ref{subsec:3.4}.

\begin{figure}[H]
  \centering
  \includegraphics[width=1\textwidth]{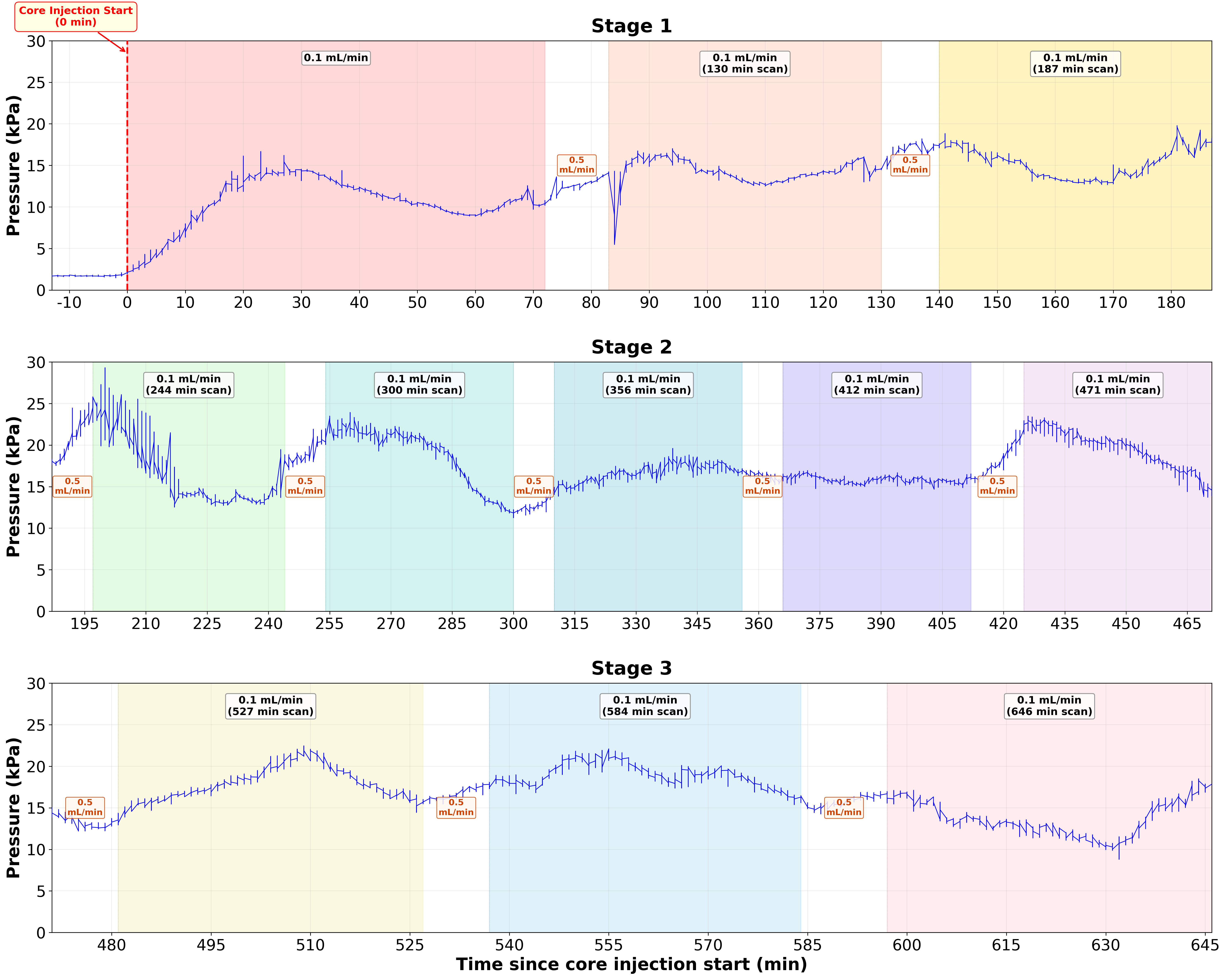}
  \caption{Pressure response over the entire experiment, displayed in
  three panels corresponding to the three evolutionary stages identified from the
  CT data. \textbf{Stage~1} (top; 0--187~min): following a transient start-up
  spike at the onset of injection ($t = 0$), the pressure has minor fluctuations, characteristic of
  an advection-dominated regime in which capillary barriers are sequentially
  overcome. \textbf{Stage~2} (middle; 187--471~min): the pressure signal
  transitions to high-amplitude oscillations
  (${\sim}\,13$--$29$~kPa), most intense during the 197--254~min sub-interval,
  reflecting repeated capillary build-up and partial breakthrough events caused
  by \ce{CO2}-swollen oil ganglia. After ${\sim}\,310$~min, both the mean
  pressure and oscillation amplitude gradually decline. \textbf{Stage~3} (bottom;
  471--646~min): the oscillation amplitude is markedly reduced relative to
  Stage~2. 
  Coloured bands denote periods of low-flow-rate
  injection (0.1~mL\,min$^{-1}$) during which CT scans were acquired; orange
  labels indicate brief high-flow-rate pulses
  (0.5~mL\,min$^{-1}$) applied between consecutive scan windows.
  }
  \label{fig:21}
\end{figure}

The pressure signal during Stage~3 (Figure~\ref{fig:21}, bottom panel) shows a reduced amplitude of oscillations compared with Stage~2, consistent with a progressive decrease in capillary resistance as throats widen and flow pathways stabilize.
\newgeometry{top=2.5cm,bottom=2.5cm}
\restoregeometry

\newgeometry{top=2.5cm,bottom=2.5cm}
\restoregeometry

\subsection{Pore-scale Structural Evolution: Topology and Oil Distribution}
\label{subsec:3.2}
\begin{figure}[H]
    \centering
    \includegraphics[width=1\textwidth]{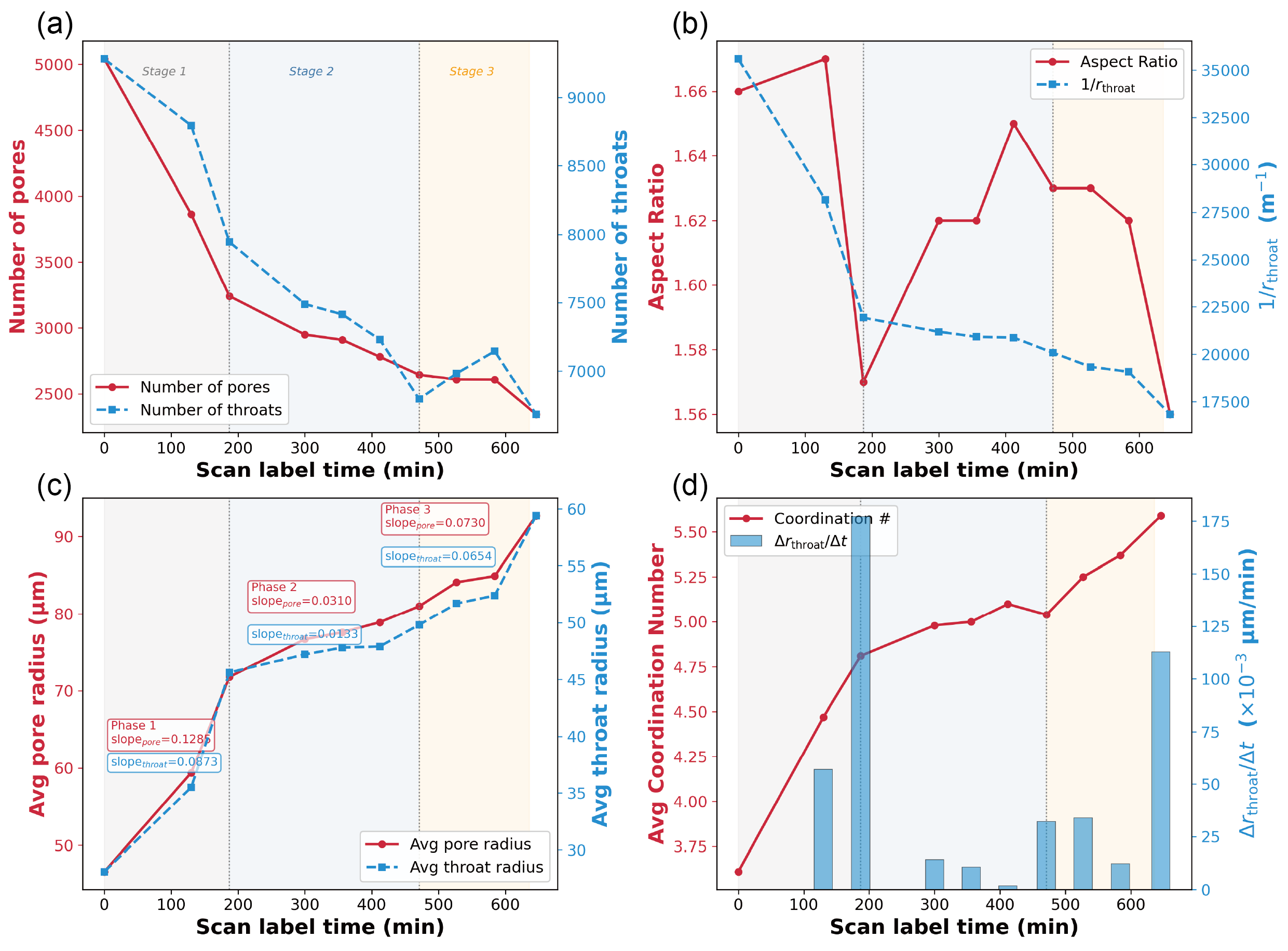}
    \caption{Temporal evolution of topological and geometric characteristics of the pore network.
    The entire evolution process is divided into three distinct stages (Stages 1, 2, and 3) based on the parameter trends, demarcated by vertical dotted lines and shaded backgrounds. 
    \textbf{(a)} The decreasing trend in the number of pores (solid red line, left y-axis) and the number of throats (dashed blue line, right y-axis) over time. 
    \textbf{(b)} Variations in the aspect ratio (solid red line, left y-axis) and the inverse of the throat radius (\(1/r_{\text{throat}}\), dashed blue line, right y-axis) over time. 
    \textbf{(c)} The increasing trend of the average pore radius (solid red line, left y-axis) and average throat radius (dashed blue line, right y-axis). The text boxes display the linear slopes (\(\mu\)m/min) of the pore and throat radii within each specific stage. 
    \textbf{(d)} Evolution of the average coordination number (solid red line, left y-axis) and the rate of change of the throat radius over time (\(\Delta r_{\text{throat}}/\Delta t\), blue bar chart, right y-axis).}
    \label{fig:3}
\end{figure}

To examine the structural changes underlying the macroscopic trends described in Section~\ref{subsec:3.1}, equivalent pore-network models were extracted from the segmented micro-CT images at each scan time. We further mapped the multiphase fluid configuration onto the extracted networks to track changes in oil occupancy.
During Stage~1, acid dissolution widens both pores (wide regions of the pore space) and throats (the restrictions between pores) that can also coalesce resulting in a decrease in the overall number of pores and throats.  The coordination number increases ($\bar{Z}$: 3.61 $\to$ 4.81; Figure~\ref{fig:3}d; Table~\ref{tab:pore_network}), consistent with enhanced connectivity.
At the same time, the aspect ratio, defined as the average ratio of the pore radius to the mean radius of connected throats, decreases modestly (from 1.66 $\to$ 1.57;
Figure~\ref{fig:3}b). A lower aspect ratio reduces the tendency for snap-off \cite{singh2022new}, which would otherwise re-trap mobilised ganglia. These structural changes---increased connectivity and reduced aspect ratio---are consistent with the rapid porosity increase and oil displacement observed at the macroscopic scale during Stage~1 (Section~\ref{subsec:3.1}).
At the Stage~1--2 transition, the rate of pore and throat enlargement slows substantially while the aspect ratio rebounds from 1.57 to 1.63 (Figure~\ref{fig:3}b; Table~\ref{tab:pore_network}). A higher aspect ratio promotes snap-off, which traps additional ganglia and blocks further throats. The coordination number $\bar{Z}$ increases only modestly during Stage~2, from 4.81 to 5.04  (Table~\ref{tab:pore_network}).
During Stage~3, the pores and throats again enlarge rapidly. A notable feature of this stage is a transient net increase of ${\sim}350$ throats over a 90\,min window (471--584\,min), the only such rebound in the entire experiment (Figure~\ref{fig:3}a). The reopened throats restore network connectivity ($\bar{Z}$: 5.04 $\to$ 5.59; Figure~\ref{fig:3}d; Table~\ref{tab:pore_network}), consistent with the renewed increase in porosity and decline in $S_{or}$ observed at the macroscopic scale (Section~\ref{subsec:3.1}).

\begin{figure}[H]
\centering
\includegraphics[width=\linewidth]{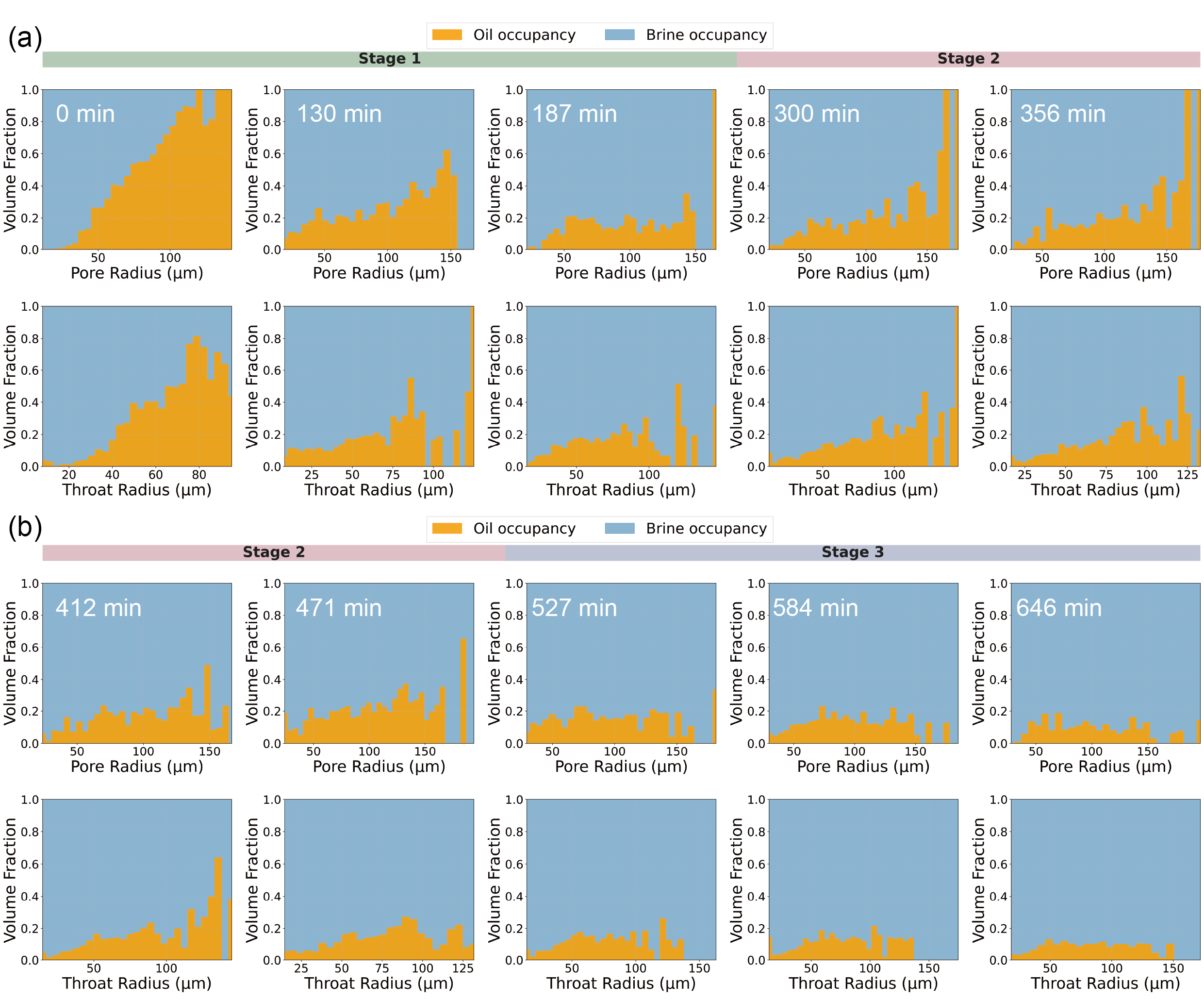}
\caption{Volume-weighted oil and brine occupancy as a function of pore radius (top rows) and throat radius (bottom rows) at successive scan times during \ce{CO2}-saturated brine injection. Orange bars denote the volume fraction occupied by the residual decane phase and blue bars denote brine occupancy within each radius bin. (a) Stage~1 and early Stage~2 (0, 130, 187, 300, 356\,min): the initial oil-saturated state at $t=0$\,min progressively evolves as dissolution enlarges pores and throats and brine displaces oil from the smaller size classes, while oil remains preferentially retained in the largest pores and throats. (b) Late Stage~2 and Stage~3 (412, 471, 527, 584, 646\,min): oil occupancy decreases across all size classes and is increasingly confined to large pores ($>100\,\mu$m), with throat-scale oil progressively cleared as dissolution-driven pore--throat enlargement reconnects the flow paths.
}
\label{fig:occupancy_volume}
\end{figure}

Pore-throat occupancy analysis (Figure~\ref{fig:occupancy_volume}) shows that the multiphase fluid configuration undergoes changes consistent with the regimes outlined previously. At 0~min, oil preferentially saturates pores and throats of intermediate to large radius (50\% of the volume in throats $>50$\,\textmu m is occupied by oil), reflecting the residual configuration established after imbibition (displacement of oil by brine). Throats with radius $>140$\,\textmu m do not yet exist at this stage; they emerge only as Stage~1 dissolution progressively widens the network. 
A pronounced shift occurs at the Stage~1$\to$2 transition. Once swelling outpaces ganglion mobilisation at 187~min (Section~\ref{subsec:3.1.2}), oil reoccupies the large throats that have just been opened by Stage~1 dissolution: the oil occupancy in throats $>140$\,\textmu m increases from 19\% at 187~min to 59\% at 300~min and reaches a peak of 73\% at 356~min. The same trend is visible at the $>100$\,\textmu m threshold (15\% $\to$ 28\% across the same interval), but is absent at the $>50$\,\textmu m threshold, where occupancy remains nearly constant—indicating that the increase in oil occupancy is concentrated in the large-throat fraction of the network, where ganglia preferentially block the brine flow. 
During the remainder of Stage~2 (356$\to$471~min), oil occupancy in the $>140$\,\textmu m class fluctuates substantially, declining to 11\% at 412~min and then increasing to 29\% at 471~min. This non-monotonic evolution reflects the dynamic competition between throat widening by dissolution and pore-space occupation by swelling ganglia, with neither process dominating; the fluctuations occur in close coordination with corresponding changes in the effective reaction rate (Section~\ref{subsec:3.3}), a coupling examined quantitatively in Section~\ref{subsec:3.4}.

During Stage~3 (471$\to$646~min), oil is rapidly displaced from the large throats: occupancy in throats $>140$\,\textmu m drops from 29\% to 0\% during 471--584~min and remains below 5\% at 646~min. The recovery is also visible at the $>100$\,\textmu m and $>50$\,\textmu m thresholds, but is most rapid and complete in the largest throats. Across the experiment, oil occupancy in the large-throat fraction of the network exhibits a non-monotonic trajectory that closely tracks the macroscopic dissolution response: rapid clearance during Stage~1, sharp rebound and dynamic fluctuation during Stage~2, and rapid evacuation during Stage~3. The mechanistic implications of this synchrony—particularly the role of pore-scale oil occupancy in modulating the effective dissolution rate observed in Section~\ref{subsec:3.3}—are examined in Section~\ref{subsec:3.4}.
\begin{table}[H]
  \centering
  \caption{Pore network parameters at different dissolution times}
  \label{tab:pore_network}
  \renewcommand{\arraystretch}{1.3}
  \resizebox{\textwidth}{!}{%
  \begin{tabular}{ccccccc}
    \toprule
    \textbf{Time} & \textbf{Number of pores} & \textbf{Number of throats} & \textbf{Average pore radius/m} & \textbf{Average throat radius/m} & \textbf{Aspect ratio} & \textbf{Average coordination number} \\
    \midrule
    0\,min   & 5040 & 9282 & $4.66\times10^{-5}$ & $2.81\times10^{-5}$ & 1.66 & 3.61 \\
    130\,min & 3863 & 8796 & $5.94\times10^{-5}$ & $3.55\times10^{-5}$ & 1.67 & 4.47 \\
    187\,min & 3241 & 7945 & $7.18\times10^{-5}$ & $4.56\times10^{-5}$ & 1.57 & 4.81 \\
    300\,min & 2949 & 7489 & $7.67\times10^{-5}$ & $4.72\times10^{-5}$ & 1.62 & 4.98 \\
    356\,min & 2909 & 7414 & $7.76\times10^{-5}$ & $4.78\times10^{-5}$ & 1.62 & 5.00 \\
    412\,min & 2781 & 7230 & $7.89\times10^{-5}$ & $4.79\times10^{-5}$ & 1.65 & 5.10 \\
    471\,min & 2645 & 6796 & $8.10\times10^{-5}$ & $4.98\times10^{-5}$ & 1.63 & 5.04 \\
    527\,min & 2609 & 6983 & $8.41\times10^{-5}$ & $5.17\times10^{-5}$ & 1.63 & 5.25 \\
    584\,min & 2608 & 7145 & $8.49\times10^{-5}$ & $5.24\times10^{-5}$ & 1.62 & 5.37 \\
    646\,min & 2345 & 6684 & $9.27\times10^{-5}$ & $5.94\times10^{-5}$ & 1.56 & 5.59 \\
    \bottomrule
  \end{tabular}%
  }
\end{table}

\subsection {Interfacial Area and Dissolution Rate}
\label{subsec:3.3}
\begin{figure}[H]
    \centering
    \includegraphics[width=1\textwidth]{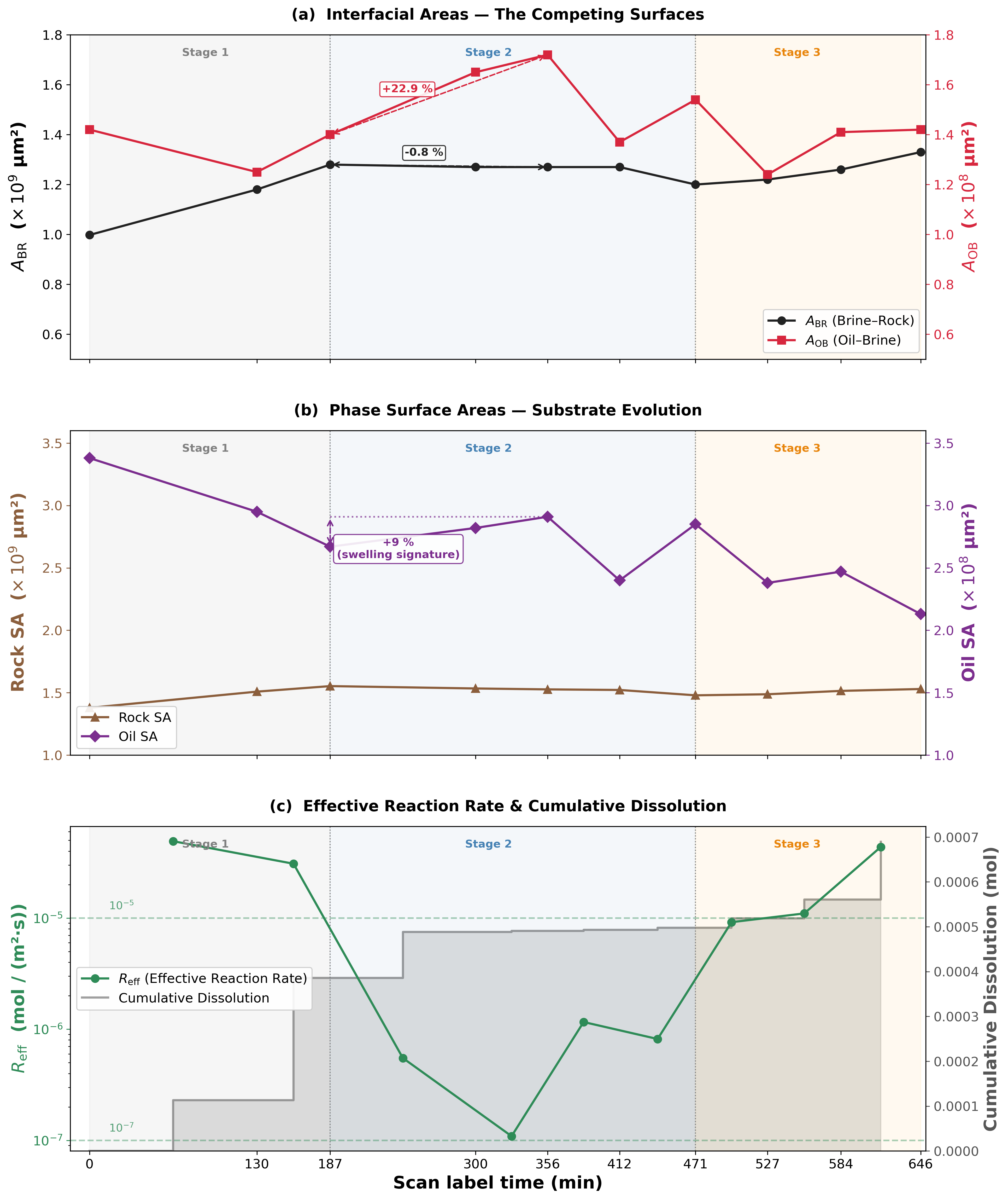}
    \caption{Interfacial areas, phase surface areas, and effective reaction rate, Eq.~(\ref{eq:reaction rate}), as a function of time across the three reaction stages. \textbf{(a)}~Brine--rock interfacial area ($A_\mathrm{BR}$, black circles, left axis) and oil--brine interfacial area ($A_\mathrm{OB}$, red squares, right axis). During Stage~2, the anti-correlation between the two areas reflects the expanding mass-transfer interface ($A_\mathrm{OB}$) displacing the reactive interface ($A_\mathrm{BR}$). \textbf{(b)}~Total rock surface area (Rock\,SA, brown triangles, left axis) and total oil surface area (Oil\,SA, purple diamonds, right axis). The $9$\,\% increase in Oil\,SA during 187--356\,min is a direct area-based signature of \ce{CO2}-induced swelling; \textbf{(c)}~Averaged effective reaction rate ($R_\mathrm{eff}$, green circles, left logarithmic axis) and cumulative moles dissolved (grey step profile, right linear axis).The effective reaction rate at \SI{65}{\minute} represents the average state from \(t = 0\) to \(t =\) \SI{130}{\minute} (\(\Delta t=65\) min), and \SI{158.5}{\minute} corresponds to the second scan performed from \(t = 130\) to \(t =\) \SI{187}{\minute}, which represents the average state at 158.5 min (\(\Delta t=45\) min), and so on for subsequent time steps.}
    \label{fig:4}
\end{figure}

Figure~\ref{fig:4} tracks the brine--rock interfacial area ($A_\mathrm{BR}$), the oil--brine interfacial area ($A_\mathrm{OB}$), the total rock and oil surface areas (Rock\,SA and Oil\,SA), the effective reaction rate ($R_\mathrm{eff}$; Eq.~(\ref{eq:reaction rate})), and cumulative dissolution over the full experimental duration.
During Stage~1, \(A_\mathrm{BR}\) rises from \(9.98\times10^{8}\) to \(1.28\times10^{9}~\mu\mathrm{m}^{2}\) (Figure~\ref{fig:4}a), as dissolution exposes fresh carbonate surface to acidic brine. Rock~SA increases concurrently (Figure~\ref{fig:4}b), reflecting growth of the resolved reactive rock surface through pore-wall exposure and corrugation at the 6\,\(\mu\)m CT resolution, together with sub-micrometre roughening indicated by SEM imaging (Appendix Figure~\ref{appendix:SEM}). 
In contrast, Oil\,SA declines steadily throughout Stage~1, consistent with effective displacement of oil ganglia. \(A_\mathrm{OB}\) first decreases and then partially recovers (Figure~\ref{fig:4}a), remaining approximately constant overall, consistent with competing effects of oil displacement and incipient \ce{CO2}-driven swelling. \(R_\mathrm{eff}\) remains high and cumulative dissolution increases steeply throughout Stage~1 (Figure~\ref{fig:4}c), indicating that expanding reactive interfaces, sustained by unobstructed advective acid supply, support uninhibited reaction.

During early Stage~2 (187--356\,min), \(A_\mathrm{OB}\) increases by \(23\%\), while \(A_\mathrm{BR}\) declines by only \(0.8\%\) (Figure~\ref{fig:4}a). This is consistent with swelling oil ganglia expanding into brine-occupied pore space. Because the sample is water-wet, a continuous brine wetting film is expected to persist between oil and rock which transfers acid to the rock surface by diffusion \cite{hirasakl1991wettability}; however, swelling can thin this film and suppress its advective exchange with bulk brine. Thus, the CT-derived \(A_\mathrm{BR}\) should be interpreted as a segmented contact area rather than the true effective reactive area: sub-voxel films may be missed, while resolved but stagnant brine--rock contacts may contribute little to dissolution. Oil\,SA increases by \(9\%\) from its Stage~1 minimum (Figure~\ref{fig:4}b), providing an area-based signature of \ce{CO2}-induced swelling. Rock\,SA remains nearly unchanged, indicating that dissolution has largely stalled. This is confirmed by the approximately two-order-of-magnitude decrease in \(R_\mathrm{eff}\) relative to Stage~1 and by the near-plateau in cumulative dissolution (Figure~\ref{fig:4}c). The strong reduction in \(R_\mathrm{eff}\) despite nearly unchanged segmented \(A_\mathrm{BR}\) indicates that rate suppression is controlled primarily by transport limitations at or near the mineral surface, rather than by loss of geometric brine--rock contact. The mechanistic interpretation combining this observation with the pore-occupancy and flow-field analyses is presented in Section~\ref{subsec:3.4}.

During late Stage~2, \(A_\mathrm{OB}\) and Oil\,SA both decline between 356 and 412\,min (Figures~\ref{fig:4}a,b), consistent with displacement of swollen oil ganglia from the imaged domain. \(R_\mathrm{eff}\) remains strongly suppressed during this interval, although a minor cumulative dissolution increment is still recorded (Figure~\ref{fig:4}c). Between 412 and 471\,min, \(A_\mathrm{OB}\) and Oil\,SA partially recover, indicating continued \ce{CO2} uptake and re-expansion of the remaining oil ganglia.

During Stage~3, \(A_\mathrm{OB}\) decreases while \(A_\mathrm{BR}\) resumes its upward trend, reaching its experimental maximum by 646\,min (Figure~\ref{fig:4}a). Oil\,SA enters a sustained decline as swollen ganglia are remobilised through the reopened throats documented in Section~\ref{subsec:3.2} (Figure~\ref{fig:4}b). \(R_\mathrm{eff}\) recovers to near its Stage~1 magnitude, and cumulative dissolution resumes a steep rise (Figure~\ref{fig:4}c).

\newgeometry{top=2.5cm,bottom=2.5cm}
\restoregeometry

\subsection{Mechanism of Rate Suppression}
\label{subsec:3.4}

\begin{figure}[H]
  \centering
  \includegraphics[width=1\linewidth]{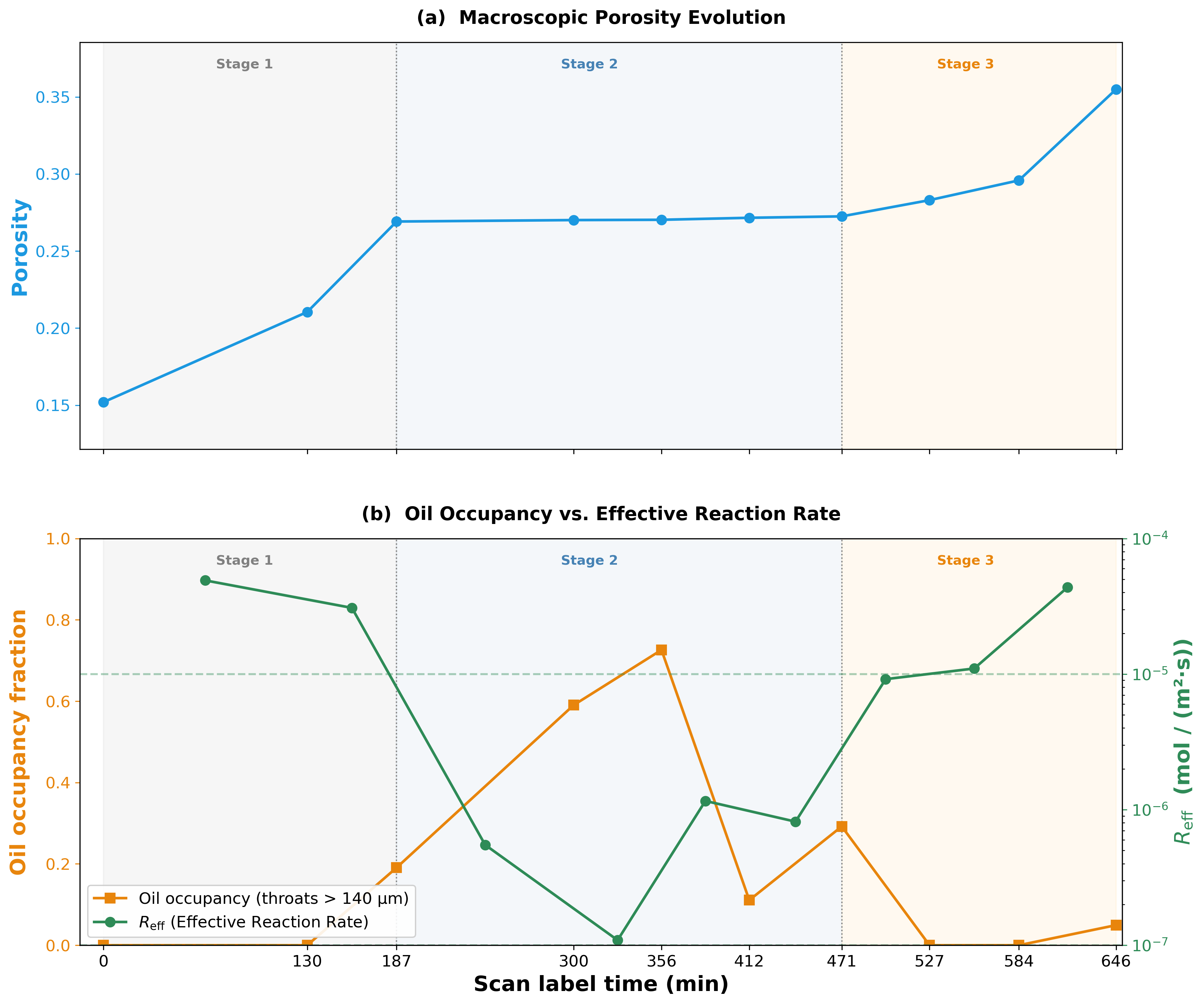}
  \caption{Co-evolution of oil occupancy in large throats and the
  effective dissolution rate throughout the experiment. Orange
  squares (left axis, linear scale) denote the volume-weighted
  fraction of throat volume occupied by the residual oil phase
  for throats with radius greater than $140$~\textmu m, extracted
  from pore-network analysis of the segmented micro-CT images.
  Green circles (right axis, logarithmic scale) denote the
  effective dissolution rate $R_{\mathrm{eff}}$
  (mol~m$^{-2}$~s$^{-1}$), computed as defined in
  Section~\ref{subsec:3.3}. Background shading marks the three-stage
  classification introduced in Section~\ref{subsec:3.1}: Stage~1
  (advection-dominated dissolution, $0$--$187$~min), Stage~2
  (suppression, $187$--$471$~min) and Stage~3 (recovery,
  $471$--$646$~min). Throats larger than $140$~\textmu m do not
  yet exist at $t=0$ and emerge progressively as Stage~1 dissolution
  widens the network. Occupancy rises sharply from $19\%$ at
  $187$~min to $73\%$ at $356$~min as $R_{\mathrm{eff}}$ collapses
  by approximately two orders of magnitude in Stage~2. In
  Stage~3 occupancy drops to below $5\%$ as $R_{\mathrm{eff}}$
  recovers toward its Stage~1 magnitude. The Pearson correlation
  between the interpolated occupancy and $\log_{10} R_{\mathrm{eff}}$
  across all nine reactive scans is $r = -0.92$.}
  \label{fig:occupancy_Reff_3phase}
\end{figure}

Sections~\ref{subsec:3.1}--\ref{subsec:3.3} provide three complementary observations that constrain the mechanism of Stage~2 rate suppression: the non-monotonic macroscopic dissolution trajectory, the preferential occupation of large throats by trapped oil, and the decoupling of the effective reaction rate from the brine--rock interfacial area. 
Figure~\ref{fig:occupancy_Reff_3phase} presents, on a common time axis, (a) the macroscopic porosity evolution and (b) the oil-occupancy fraction in large throats ($>140\,\mu\mathrm{m}$ 
radius) together with the effective reaction rate $R_\mathrm{eff}$. Oil occupancy and $R_\mathrm{eff}$ (Figure~\ref{fig:occupancy_Reff_3phase}b) exhibit closely coupled, non-monotonic co-evolution through the same three phases identified from the macroscopic trajectory (Figure~\ref{fig:occupancy_Reff_3phase}a). Stage~1 is characterized by absent or low occupancy in the large-throat class (which emerges only late in Stage~1) and high $R_\mathrm{eff}$ ($\sim$$10^{-5}$~mol\,m$^{-2}$\,s$^{-1}$). 
At the Stage~1--Stage~2 transition, occupancy rises sharply, from 19\% at 187~min to 59\% at 300~min and 73\% at 356~min, coinciding with a collapse of \(R_\mathrm{eff}\) by about two orders of magnitude to \(1.1\times10^{-7}\,\mathrm{mol\,m^{-2}\,s^{-1}}\) at 328~min. During Stage~2, both quantities continue to fluctuate in apparent coordination: a transient drop in occupancy to 11\% at 412~min is accompanied by a partial recovery in $R_\mathrm{eff}$ (an order-of-magnitude rebound), followed by a renewed occupancy increase to 29\% at 471~min and a corresponding decline in $R_\mathrm{eff}$. 
In Stage~3, occupancy drops to near zero and $R_\mathrm{eff}$ recovers toward its Stage~1 magnitude. The Pearson correlation between interpolated occupancy and $\log_{10} R_\mathrm{eff}$ across all nine reactive scans is $r = -0.92$, indicating strong correlation between oil being in large throats and reactive suppression.

The synchrony observed above could in principle reflect either a geometric or chemical control: oil in throats might suppress reaction by reorganising the advective network (geometric), or by absorbing \ce{CO2} from the bulk brine and reducing acid availability downstream (chemical). Mass-balance considerations rule out the chemical pathway as the primary driver. The cumulative \ce{CO2} delivered during Stage~2 ($\sim$48~mmol) far exceeds both the saturation capacity of the residual decane phase ($\leq$0.54~mmol) and the \ce{CO2} consumed by reaction ($\sim$0.04~mmol). The bulk brine therefore remains close to its injection acidity except during the early partition transient, and the cumulative chemical effect of partition is bounded well below the \ce{H+} deficit implied by the rate suppression.

\begin{figure}[H]
\centering
\includegraphics[width=\textwidth]{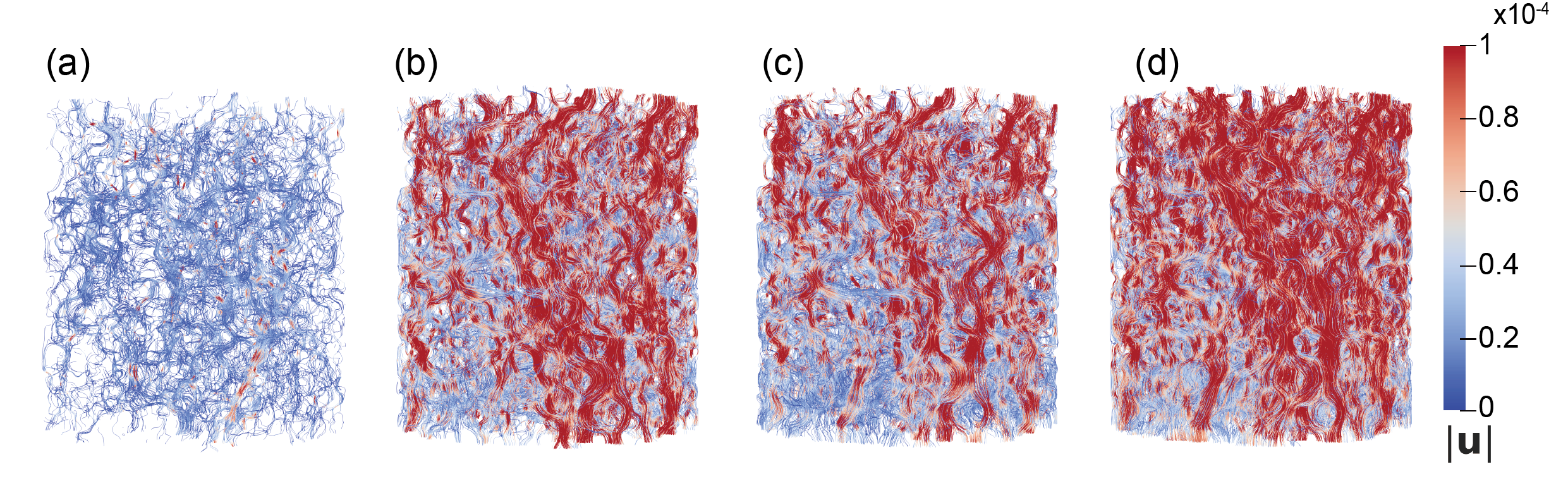}
\caption{Pore-scale streamlines coloured by velocity magnitude 
$|\mathbf{u}|$ (m\,s$^{-1}$) at four representative timepoints. 
(a)~$t = 0$~min (start of Stage~1): flow distributes relatively 
uniformly with high-velocity streamlines (red) appearing only 
sporadically. (b)~$t = 300$~min (start of Stage~2): Stage~1 
dissolution has restructured the flow field into a markedly 
heterogeneous distribution, with elevated velocities along 
preferential flow paths and stagnant zones in the surrounding 
pore space. (c)~$t = 471$~min (end of Stage~2): the heterogeneous 
flow distribution persists with little overall change. 
(d)~$t = 646$~min (end of Stage~3): the preferential flow paths 
further intensify as oil evacuation from large throats restores 
advective flux along these paths.}
\label{fig:streamline}
\end{figure}

Pore-scale flow simulations to solve Eqs.~(1,2) were performed on the segmented micro-CT volumes within the \texttt{OpenFOAM} framework (Section~\ref{sec:2.5}), with residual oil treated as immobile; the resulting streamlines (Figure~\ref{fig:streamline}) reveal the geometric pathway reorganisation between experimental states. 
At $t = 0$ early in Stage~1, flow is distributed relatively uniformly across the sample. By 300~min, Stage~1 dissolution has restructured the flow field into a markedly heterogeneous distribution in which high-velocity streamlines concentrated along preferential flow paths while the surrounding pore space is occupied by low-velocity streamlines and stagnant zones. 
This heterogeneous distribution persists with little overall change through 471~min, indicating that during the rate-suppression interval the trapped-oil configuration locks the flow field into a quasi-stationary state. By 646~min, at the end of Stage~3, the preferential flow paths have further intensified as oil displacement from large throats restores advective flux. Throughout Stages~2 and~3, mineral surfaces outside the high-velocity flow paths are contacted only by stagnant or slowly recirculating brine.
Together, these three observations—closely coupled non-monotonic co-evolution of occupancy and $R_\mathrm{eff}$, mass-balance constraints, and heterogeneous flow with stagnant zones—motivate the mechanistic interpretation developed below.

Within Stage~2, where swelling outpaces ganglion mobilisation (Section~\ref{subsec:3.1.2}), three spatially coupled effects—initiated or amplified by \ce{CO2}-induced ganglion swelling—account for the observed rate suppression. 
Of these, the geometric reorganisation of advective access (i) is dominant, with chemical (ii) and kinematic (iii) effects acting as subordinate contributions.

\textbf{(i) Geometric severance of advective access.} Oil persistently occupies the large-throat fraction of the network (Section~\ref{subsec:3.2}), constricting the preferential flow paths visualised in Figure~\ref{fig:streamline}. Brine must still traverse the network, but the lateral advective exchange between these preferential flow paths and surrounding pore clusters is impeded as ganglia partially obstruct connecting throats. We hypothesize that mineral surfaces in the surrounding regions---which constitute much of the segmented $A_\mathrm{BR}$---lose effective advective contact with fresh acidic brine; local \ce{H+} is consumed, dissolution products (\ce{Ca^{2+}}, \ce{HCO3^{-}}) accumulate, and the fluid drifts toward equilibrium with the mineral surface, suppressing the forward reaction~\cite{myint2015thin}. Where ganglia swell directly against pore walls along these flow paths, the persistent wetting brine film~\cite{hirasakl1991wettability} also loses advective replenishment, producing the analogous film-confined regime described by Nishiyama and Yokoyama~\cite{nishiyama2021water}. This loss of advective contact across most of $A_\mathrm{BR}$ accounts for the decoupling of $R_\mathrm{eff}$ from $A_\mathrm{BR}$ documented in Section~\ref{subsec:3.3}.

\textbf{(ii) Subordinate chemical effect from \ce{CO2} partitioning.} Although the global chemical effect of \ce{CO2} partitioning into the oil phase is bounded by mass balance, we suggest that steep concentration gradients at the oil--brine interface produce a localised reduction in \ce{H+} activity. \ce{CO2} partitions into the trapped oil phase and shifts the coupled carbonate equilibria leftward. The related equations are \citep{plummer1978kinetics, peng2015kinetics}:

\begin{align}
\ce{CO2(aq) + H2O &<=> H2CO3 <=> H+ + HCO3-} 
  \quad  \label{eq:carbonic}\\
\ce{CaCO3 + H+ &<=> Ca^{2+} + HCO3-} 
  \quad  \label{eq:calcite-dissolution}\\
\ce{CO2(aq) &<=> CO2(oil)} 
  \quad  \label{eq:co2-partition}
\end{align}

As \ce{CO2(aq)} is drawn into the oil,  Reaction~\eqref{eq:carbonic} shifts leftward, locally lowering \ce{H+} activity in the brine adjacent to ganglia and driving Reaction~\eqref{eq:calcite-dissolution} towards equilibrium at nearby mineral surfaces.

\textbf{(iii) Reduced mass transfer in widened Stage~1 flow paths.} Where Stage~1 dissolution widens connected flow paths, the same imposed flow is distributed over a larger local cross-section, reducing the near-wall velocity gradient and hence the surface-normal mass transfer. This may lead to a weaker \ce{H+} resupply and product removal at mineral surfaces beyond the swollen ganglia.

This proposed three-effect picture sits within the broader swelling--displacement competition introduced in 
Section~\ref{subsec:3.1}. Because phase-equilibrium considerations imply a sustained driving force for \ce{CO2} partitioning into the oil phase throughout the experiment (Section~\ref{subsec:3.1.1}), ganglion swelling occurs continuously across all three stages. The effective reaction rate $R_\mathrm{eff}$ reflects the balance between two competing processes: the rate at which swelling accumulates oil within large throats, and the rate at which advective forces mobilise oil before such accumulation.
In Stage~1, throat widening removes ganglia faster than swelling can stabilise them in large throats, leaving advective access to fresh acid intact. In Stage~2, swelling outpaces mobilisation: ganglia accumulate in large throats faster than they can be displaced, reorganising the advective field. In Stage~3, dissolution-driven pore merging enlarges flow paths and creates bypass routes around occupied throats, restoring advective contact with downstream mineral surfaces.

In our previous work~\cite{ma2026pore} a Ketton sample initially saturated with brine and decane was injected with \ce{CO2}-saturated brine at \(0.05~\mathrm{mL\,min^{-1}}\) for \(315~\mathrm{min}\), after which the flow rate was increased to \(0.5~\mathrm{mL\,min^{-1}}\) and maintained for a further \(225~\mathrm{min}\). 
The high-rate stage was accompanied by both an increase in dissolution rate and a continued increase in oil-phase saturation, where oil from downstream regions transported into highly conductive channels and accumulated as large ganglia, as well as continued \ce{CO2}-induced swelling. 
This behaviour is consistent with the mechanistic hypothesis presented here: under sufficient advective forcing, mobilisation outpaces swelling, and the reaction rate recovers despite ongoing partitioning of \ce{CO2} into the oil. 

\subsection{Implications for Reactive-Transport Modelling}
\label{subsec:3.5}
The decoupling of $R_\mathrm{eff}$ from $A_\mathrm{BR}$ documented in Section~\ref{subsec:3.3} indicates that interfacial area alone is not a sufficient determinant for local dissolution rate. Local flow velocity and the dimensionless groups built from it—such as the Péclet and Damköhler numbers—are similarly insufficient when the trapped-phase configuration redistributes advective access in ways not captured by their bulk values. Although pore-scale direct numerical simulations with thermodynamically consistent interphase transfer could, in principle, reproduce such behaviour~\cite{maes2021geochemfoam, soulaine2018pore}, the specific effect considered here---whereby a trapped non-wetting phase along the flow path reduces the advective access of acidic brine to downstream mineral surfaces---has not been identified or quantified in previous multiphase reactive-transport studies. 
Upscaled models based only on local closures---whether geometric, kinematic, or statistical---may therefore fail to capture the path-dependent behaviour observed in our experiments. Predictive modelling of such systems will likely require explicit representation of trapped-phase configuration and its evolution.

\subsection{Applicability and Limitations}
\label{subsec:3.6}
The principal results presented above rest on a single experiment conducted on one water-wet Ketton limestone sample. While the internal consistency among the porosity, pore-network, interfacial-area, and streamline datasets supports the interpretations offered, and the consistency check from our previous work (Section~\ref{subsec:3.4}) further constrains the interpretation, several factors may prevent direct generalisation. 

Ketton limestone possesses a well-connected, bimodal pore structure dominated by intergranular macropores; carbonates with more heterogeneous pore architectures may exhibit different patterns of throat blockage and flow-pathway reorganisation. 

Decane was chosen as the oil phase for its well-characterised properties but lacks the polar and asphaltenic components of reservoir crude oils that can modify wettability, interfacial tension, and \ce{CO2} solubility. The behaviour may be different when using crude oils that will affect the rock wettability.

The localised \ce{H+} weakening pathway (mechanism~(ii) of Section~\ref{subsec:3.4}) is inferred rather than measured; future experiments incorporating effluent or in situ chemical analysis could test this pathway. Finally, the alternating injection protocol introduces periodic perturbations whose cumulative effect has not been isolated from the intrinsic pore-system dynamics.

Therefore future work could focus on a systematic investigation of wettability, trapped-phase composition, and pore-structure heterogeneity. The stage-boundary times and magnitude of rate suppression should be regarded as condition-specific rather than universal. The primary findings are the qualitative existence of a three-regime, threshold-bounded dissolution trajectory and the dominance of geometric advective severance over bulk acid depletion as the suppression mechanism.

\section{Conclusions}
Time-resolved micro-CT imaging, pore-network extraction, interfacial-area analysis, and continuous pressure monitoring were combined to track the injection of \ce{CO2}-saturated brine into a water-wet Ketton limestone sample containing residual hydrocarbons. 
This integrated approach links pore-scale dissolution, hydrodynamic response, oil redistribution/swelling, and interfacial access during reactive flow in the presence of a non-wetting phase.

In our experiment, the dissolution rate was non-monotonic in time and proceeded through three distinct stages, separated by transitions at approximately \SI{187}{\minute} and \SI{471}{\minute} under the specific conditions investigated. During Stage~1, porosity increased steadily as oil ganglia were displaced and pore-network connectivity improved. 
During Stage~2, porosity plateaued, the hydrocarbon phase exhibited a net volume increase as continued \ce{CO2} partitioning outpaced ganglion mobilisation, and the effective reaction rate decreased by approximately two orders of magnitude against an essentially unchanged brine--rock interfacial area; injection pressure exhibited high-amplitude oscillations consistent with repeated capillary build-up and partial breakthrough events. 
During Stage~3, porosity increased again rapidly and the effective reaction rate recovered towards its Stage~1 magnitude.

Pore-network occupancy analysis showed that oil preferentially occupied the largest throats during Stage~2, and that the oil-occupancy fraction in throats larger than \SI{140}{\micro\metre} varied inversely with the effective reaction rate across the experiment (Pearson coefficient \(r=-0.92\) between occupancy and the logarithm of the effective reaction rate \(\log_{10} R_{\mathrm{eff}}\)). These observations indicate a shifting balance between dissolution-induced throat widening and oil-ganglion swelling. We therefore attribute the Stage 2 rate suppression to three coupled effects: swollen ganglia obstructed large throats and severed advective access to parts of the mineral surface, \ce{CO2} partitioning into adjacent oil locally weakened aqueous acidity, and reduced near-wall mass transfer in widened flow paths further limited \ce{H+} resupply to shielded surfaces.

The transitions at \SI{187}{\minute} and \SI{471}{\minute} mark changes in the balance between hydrocarbon swelling and ganglion mobilisation, producing a self-amplifying loss of advective access in Stage~2 and its gradual release in Stage~3.
Such history-dependent coupling between multiphase displacement, oil swelling, and advective access to the reactive interface cannot be represented by treating the trapped phase as static or inert.

Two aspects of these results carry broader implications for subsurface reactive transport. 
First, the decoupling of segmented brine--rock interfacial area from $R_\mathrm{eff}$ during Stage 2 shows that considering the total reactive interface alone is insufficient: the interface must also remain advectively connected. Swollen and redistributed oil ganglia can partition the mineral surface into advectively connected and advectively isolated domains, so upscaled models based only on local interfacial area, velocity, or dimensionless groups may miss this path-dependent loss of advective access.
Second, the reversible onset and release of dissolution suppression imply that near-wellbore porosity and permeability evolution can be strongly history dependent, so short-term observations may not extrapolate reliably to longer injection periods.

These findings are directly relevant to carbonate reservoirs targeted for geological \ce{CO2} storage, particularly depleted hydrocarbon reservoirs where residual oil is present and where predictions of injectivity, reactive surface evolution, and long-term containment depend on accurately representing \ce{CO2}--brine--oil--rock coupling. The pore-scale feedback observed here---\ce{CO2} uptake and swelling/redistribution of trapped oil reorganizing advective access and thereby suppressing or releasing dissolution---merits consideration in near-wellbore reactive-transport models.

\newgeometry{top=2.5cm,bottom=2.5cm}
\section{Appendix}
\renewcommand{\thefigure}{A\arabic{figure}}
\setcounter{figure}{0}

\begin{figure}[H]
    \centering
    \includegraphics[width=1\textwidth]{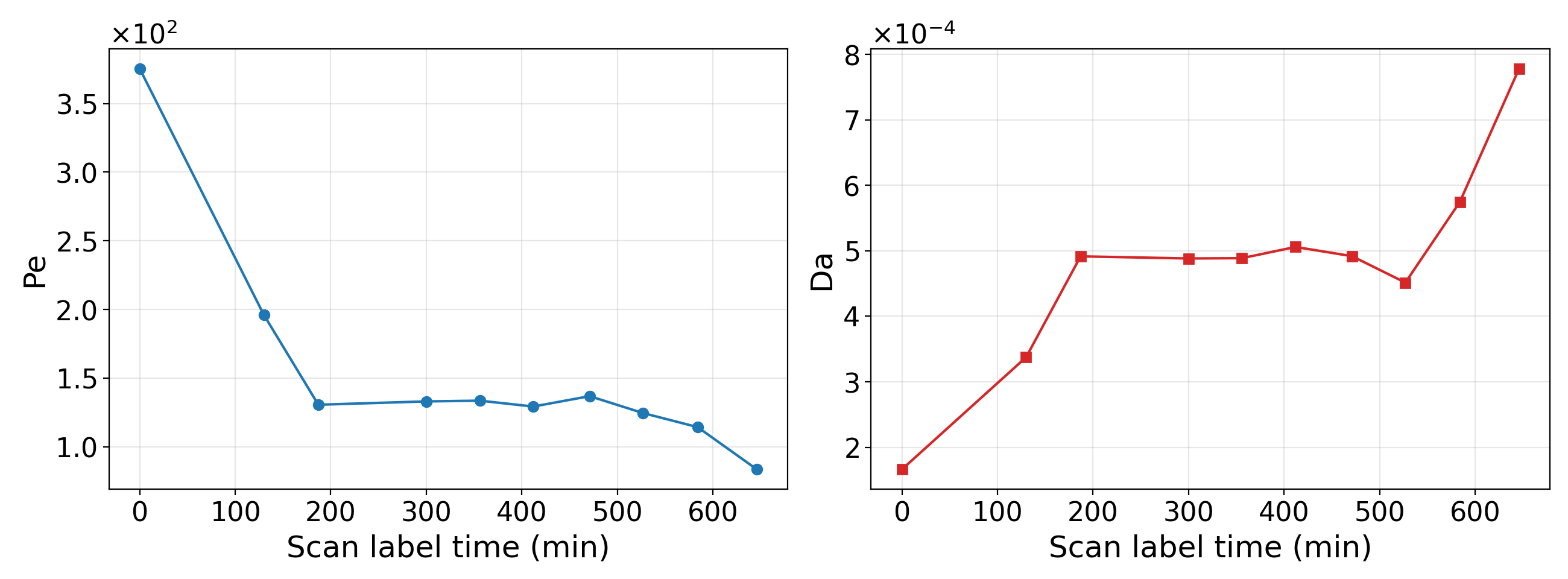}
    \caption{P\'eclet ($Pe$, Eq.~(\ref{Pe}), left) and Damk\"ohler ($Da$, Eq.~(\ref{Da}) right) numbers as a function of time, evaluated at the baseline injection rate of 0.1~mL\,min$^{-1}$. $Pe$ decreases from $375$ at $t=0$ to $83$ at 646~min, while $Da$ increases from $1.7\times10^{-4}$ to $7.7\times10^{-4}$ over the same interval, both reflecting the progressive enlargement of pore-throat radii and reduction in bulk pore velocity as dissolution proceeds.}
    \label{appendix:PeDa}
\end{figure}

\begin{figure}[H]
    \centering
    \includegraphics[width=1\textwidth]{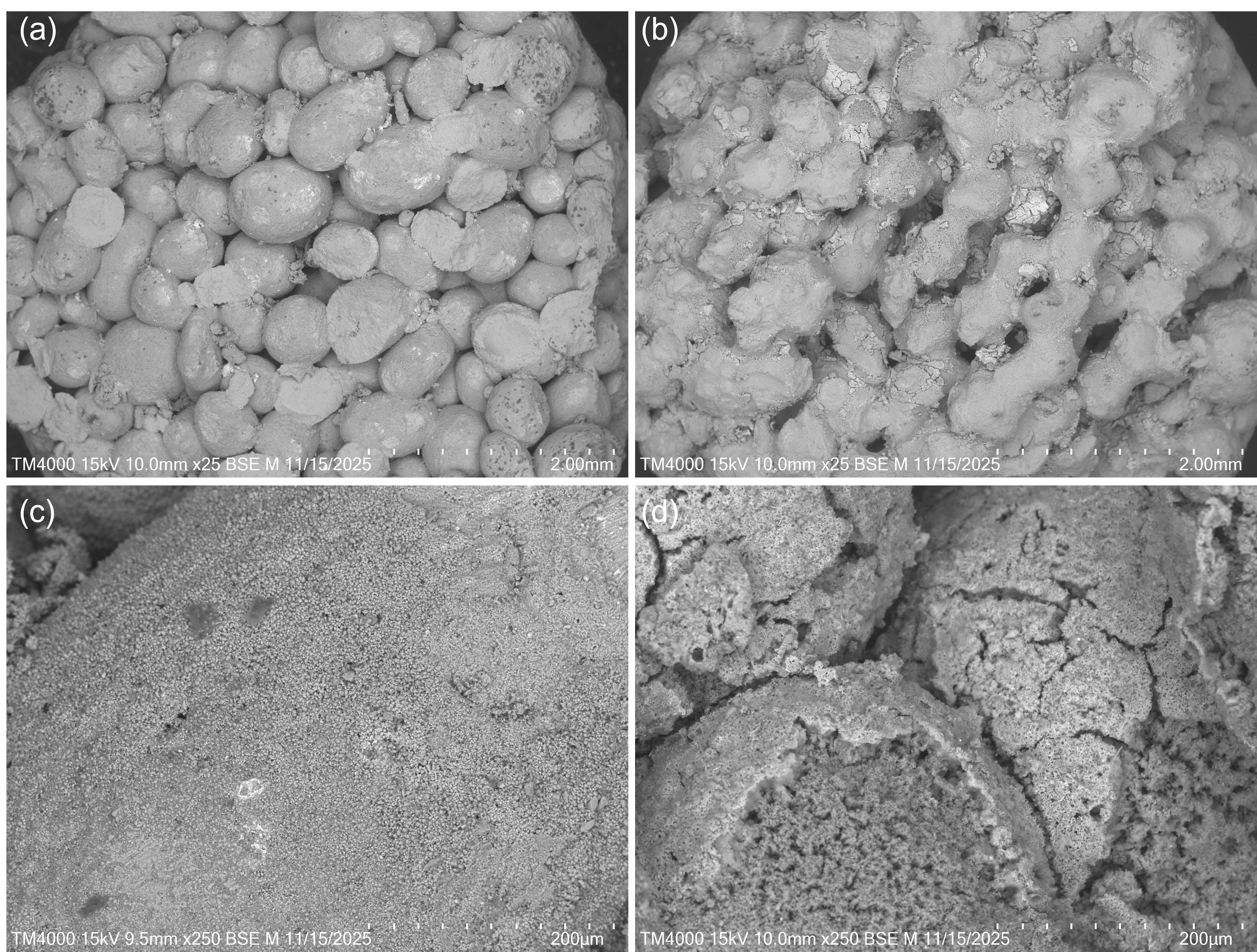}
    \caption{BSE-SEM images of rock samples before (a, c) and after (b, d) reaction  at ×25 (a, b; scale bar: 2\,mm) and ×250 (c, d; scale bar: 200\,\textmu m) magnification. Pre-reaction samples (a, c) were taken from adjacent locations on the same core. Acid exposure produces grain-scale disaggregation and sub-micrometre surface roughening, consistent with the net increase in reactive surface area observed during injection.}
    \label{appendix:SEM}
\end{figure}
\restoregeometry
\section*{List of parameters}

\begin{table}[H]
    \centering
    \caption{List of parameters, symbols, and units used in simulations and calculations.}
    \label{tab:simulation_parameters}
    \renewcommand{\arraystretch}{1.2}
    \begin{tabular}{l c c}
        \toprule
        \textbf{Parameter} & \textbf{Symbol} & \textbf{Unit} \\
        \midrule
        \multicolumn{3}{l}{\textit{Geometry and Grid}} \\
        Sample Lengths & $L_y, L_z$ & $\mathrm{m}$ \\
        Cross-sectional Area & $A_x$ & $\mathrm{m}^2$ \\
        Porosity & $\phi$ & - \\
        Unresolved Porosity & $\phi_{\text{unresolved}}$ & - \\
        Specific Surface Area & $S$ & $\mathrm{m}^{-1}$ \\
        Characteristic Length & $L_c$ & $\mathrm{m}$ \\
        
        \midrule
        \multicolumn{3}{l}{\textit{Fluid Properties and Flow}} \\
        Fluid Density & $\rho$ & $\mathrm{kg/m}^3$ \\
        Fluid Viscosity & $\mu$ & $\mathrm{Pa \cdot s}$ \\
        Pressure & $p$ & $\mathrm{Pa}$ \\
        Pressure Gradient & $\nabla p$ & $\mathrm{Pa/m}$ \\
        Velocity Vector & $\mathbf{u}$ & $\mathrm{m/s}$ \\
        Flow Rate & $Q$ & $\mathrm{m}^3/\mathrm{s}$ \\
        Darcy Velocity & $q$ & $\mathrm{m/s}$ \\
        Average Pore Velocity & $u_{\text{av}}$ & $\mathrm{m/s}$ \\
        Permeability & $K$ & $\mathrm{m}^2$ \\
        Tortuosity & $\tau$ & - \\
        
        \midrule
        \multicolumn{3}{l}{\textit{Reactive Transport}} \\
        Péclet Number & $\text{Pe}$ & - \\
        Damköhler Number & $\text{Da}$ & - \\
        Molecular Diffusion Coefficient & $D_m$ & $\mathrm{m}^2/\mathrm{s}$ \\
        Reaction Rate Constant & $k$ & $\mathrm{s}^{-1}$ \\
        Mineral Reaction Rate & $r$ & $\mathrm{mol \cdot m^{-2} \cdot s^{-1}}$ \\
        Effective Reaction Rate & $r_{\text{eff}}$ & $\mathrm{mol \cdot m^{-2} \cdot s^{-1}}$ \\
        Mineral Density & $\rho_{\text{mineral}}$ & $\mathrm{kg/m}^3$ \\
        Molecular Mass & $M_{\text{mineral}}$ & $\mathrm{kg/mol}$ \\
        Time Interval & $\Delta t$ & $\mathrm{s}$ \\
        Porosity Change & $\Delta \phi_{\text{CT}}$ & - \\
        \bottomrule
    \end{tabular}
\end{table}

\section*{Acknowledgements}

QM gratefully acknowledges Resource Geophysics Academy, Imperial College London for financial support. The authors also extend their sincere gratitude to Vincenzo Cunsolo for his invaluable assistance in conducting the experiments.

\bibliographystyle{elsarticle-num}
\bibliography{main} 
\end{document}